\documentclass[conference]{IEEEtran}
\usepackage{cite}
\usepackage{textcomp}
\usepackage{xcolor}
\usepackage{color}
\usepackage{colortbl}
\usepackage{fancyhdr}
\usepackage[hyphens]{url}
\usepackage{balance} 
\usepackage{booktabs} 
\usepackage{amsmath,amssymb,amsfonts}
\usepackage{mathtools, cases}
\usepackage{algorithm}
\usepackage{algpseudocode}
\usepackage{algorithmicx}
\usepackage{booktabs}
\usepackage{graphicx}
\usepackage{multirow}
\usepackage{multicol} 
\usepackage{comment}
\usepackage[colorinlistoftodos,prependcaption]{todonotes}
\usepackage{subcaption}
\usepackage[hidelinks]{hyperref}
\usepackage{booktabs, dcolumn}
\usepackage{soul}
\usepackage{enumitem}
\usepackage{bbding}
\usepackage{pifont}

\usepackage{dblfloatfix}
\usepackage{threeparttable}
\usepackage{bm}
\usepackage{overpic} 
\newcolumntype{d}[1]{D{.}{.}{#1}}
\usepackage{listings}

\usepackage{lipsum} 
\usepackage{xargs} 
\newlist{balditemize}{itemize}{1}
\setlist[balditemize]{label=, wide=\parindent, labelsep*=0pt, leftmargin=*, topsep=0pt, itemsep=0pt}

\usepackage[skip=4pt]{caption} 
\newcommand{\blue}[1]{\textcolor{black}{#1}}

\usepackage[utf8]{inputenc}
\usepackage{listings}
\usepackage{xcolor}

\definecolor{codegreen}{rgb}{0,0.6,0}
\definecolor{codegray}{rgb}{0.5,0.5,0.5}
\definecolor{codepurple}{rgb}{0.58,0,0.82}
\definecolor{backcolour}{rgb}{0.95,0.95,0.92}
\lstdefinestyle{mystyle}{
  backgroundcolor=\color{white},
  xleftmargin = 0.2cm,
  commentstyle=\color{codegreen},
  keywordstyle=\color{magenta},
  numberstyle=\scriptsize\color{codegray},
  stringstyle=\color{codepurple},
  basicstyle=\ttfamily\scriptsize,
  breakatwhitespace=false,         
  breaklines=true,                 
  captionpos=b,                    
  keepspaces=false,                 
  numbers=left,                    
  numbersep=5pt,                  
  showspaces=false,                
  showstringspaces=false,
  showtabs=false,                  
  tabsize=2, 
  lineskip=-1ex, 
  escapechar={|}
}
\usepackage{tikz}

\definecolor{gray}{HTML}{AEAEAE}

\def\BibTeX{{\rm B\kern-.05em{\sc i\kern-.025em b}\kern-.08em
    T\kern-.1667em\lower.7ex\hbox{E}\kern-.125emX}}

\title{ReGraph: Scaling Graph Processing on HBM-enabled FPGAs with Heterogeneous Pipelines} 

\author{Xinyu Chen$^{1}$, Yao Chen$^{2}$, Feng Cheng$^{3}$, Hongshi Tan$^{1}$,  Bingsheng He$^{1}$, Weng-Fai Wong$^{1}$ \vspace{1mm}
\\
$^{1}$National University of Singapore,
$^{2}$Advanced Digital Sciences Center,
$^{3}$City University of Hong Kong
}    

\begin{document}
\maketitle
\thispagestyle{plain}
\pagestyle{plain}


\begin{abstract}
The use of FPGAs for efficient graph processing has attracted significant interest. Recent memory subsystem upgrades including the introduction of HBM in FPGAs promise to further alleviate memory bottlenecks. However, modern multi-channel HBM requires much more processing pipelines to fully utilize its bandwidth potential.
Existing designs do not scale well, resulting in underutilization of the HBM facilities even when all other resources are fully consumed.

In this paper, we re-examined the graph processing workloads and found much diversity in processing. 
We also found that the diverse workloads can be easily classified into two types, namely dense and sparse partitions. 
This motivates us to propose a resource-efficient heterogeneous pipeline architecture.
Our heterogeneous architecture comprises of two types of pipelines: Little pipelines to process dense partitions with good locality and Big pipelines to process sparse partitions with extremely poor locality. 
Unlike traditional monolithic pipeline designs, the heterogeneous pipelines are tailored for more specific memory access patterns, and hence are more lightweight, allowing the architecture to scale up more effectively with limited resources.
In addition, we propose a model-guided task scheduling method that schedules partitions to the right pipeline types, generates the most efficient pipeline combination and balances workloads.
Furthermore, we develop an automated open-source framework, called ReGraph\footnote{ReGraph is open-sourced at https://anonymous.4open.science/r/ReGraph/.}, which automates the entire development process. 
ReGraph outperforms state-of-the-art FPGA accelerators by up to 5.9$\times$ in terms of performance and 12$\times$ in terms of resource efficiency. 

\end{abstract}

\section{Introduction}\label{sec:introduction}
Graphs are de facto data structures to represent the relationships between entities in many application domains such as social networks, genomics, and machine learning~\cite{GraphLab, Graph500}. 
As a result, efficient graph processing is becoming increasingly important, especially as the amount of graph data grows~\cite{sahu2017ubiquity}.
Allowing efficient customization on the hardware logic to computation/memory access patterns, FPGA usually delivers better memory efficiency and energy efficiency than CPUs/GPUs~\cite{Zhou2018CF,Zhou2016,ForeGraph,graphgen,zhou2019hitgraph,chen2019fly,chen2021thundergp,isca2021,ruoshi2019improving}.
Furthermore, high-level synthesis (HLS) that translates kernels written in high-level languages to low-level RTL modules alleviates the poor programmability issue of FPGAs, providing high usability to efficient graph processing systems~\cite{guo2021autobridge,HLSsurvey2016,chen2021thundergp}.

While graph processing is data access intensive, recent High Bandwidth Memory (HBM) enabled-FPGAs bring tremendous performance potentials. 
Graph processing explores the irregular structure of a graph rather than performing large numbers of computations, resulting in poor data locality and high communication to computation ratio~\cite{graphchallenges,zhu2016gemini}.
Memory bandwidth is therefore the major bottleneck to the system performance. 
Recent FPGAs have started integrating HBM to meet the demand for the ever-increasing memory bandwidth of data center applications~\cite{kara2020high,choi2021hbm}. 
For example, Alveo U280~\cite{xilinxu280} equipped with 32 HBM channels can deliver up to 460 GB/s of peak memory bandwidth, which is a sixfold increase over the latest FPGA platform with four DRAM channels (Alveo U250~\cite{xilinxu250}). 
The largely increased bandwidth from more memory channels offers to satiate much more graph processing pipelines~\cite{zhou2019hitgraph,chen2021thundergp}, hence boosting performance.

\begin{table}[t!]
\centering
\caption{Estimation of resource utilization of existing designs with increasing the number of memory channels (\#CH). }
\label{table:resource_projection}
\setlength{\tabcolsep}{3pt} 
\renewcommand{\arraystretch}{1} 
\resizebox{0.49\textwidth}{!}{%
\begin{tabular}{l|c|rcccc}
\toprule
\begin{tabular}[c]{@{}l@{}}Existing\\ Designs\end{tabular} & \multicolumn{1}{l|}{\begin{tabular}[c]{@{}l@{}}Resource\\ Type\end{tabular}} & \multicolumn{1}{c}{\begin{tabular}[c]{@{}c@{}}1 CH\\ (14 GB/s)\end{tabular}} & \multicolumn{1}{c}{\begin{tabular}[c]{@{}c@{}}4 CH\\ (58 GB/s)\end{tabular}} & \multicolumn{1}{c}{\begin{tabular}[c]{@{}c@{}}8 CH\\ (115 GB/s)\end{tabular}} & \multicolumn{1}{c}{\begin{tabular}[c]{@{}c@{}}16 CH\\ (230 GB/s)\end{tabular}} & \multicolumn{1}{c}{\begin{tabular}[c]{@{}c@{}}32 CH\\ (460 GB/s)\end{tabular}} \\ \midrule

HitGraph~\cite{zhou2019hitgraph} & LUT  & \cellcolor[rgb]{ .988,  .988,  1} *16.9\% & \cellcolor[rgb]{ .988,  .953,  .965}*68.1\% & \cellcolor[rgb]{ .988,  .906,  .914}136.2\% & \cellcolor[rgb]{ .984,  .808,  .816}272.4\% & \cellcolor[rgb]{ .98,  .612,  .62}544.8\% \\

FabGraph~\cite{ruoshi2019improving}  & LUT &\cellcolor[rgb]{ .988,  .984,  .996} *25.5\% & \cellcolor[rgb]{ .988,  .929,  .941}102.1\% & \cellcolor[rgb]{ .988,  .855,  .867}204.2\% & \cellcolor[rgb]{ .984,  .71,  .718}408.5\% & \cellcolor[rgb]{ .973,  .412,  .42}817.0\% \\

ISCA'21~\cite{isca2021}  & LUT & \cellcolor[rgb]{ .988,  .988,  1} 18.6\% & \cellcolor[rgb]{ .988,  .949,  .961}*74.2\% & \cellcolor[rgb]{ .988,  .894,  .906}148.4\% & \cellcolor[rgb]{ .984,  .788,  .8}296.8\% & \cellcolor[rgb]{ .98,  .576,  .584}593.6\% \\

ThunderGP~\cite{chen2021thundergp}  & CLB  &\cellcolor[rgb]{ .988,  .988,  1} 21.3\% & \cellcolor[rgb]{ .988,  .941,  .953}*85.3\% & \cellcolor[rgb]{ .988,  .878,  .89}170.6\% & \cellcolor[rgb]{ .984,  .757,  .769}341.2\% & \cellcolor[rgb]{ .976,  .51,  .518}682.4\% \\ \bottomrule
\end{tabular}
}
\begin{tablenotes}[flushleft]
\item {*Numbers are obtained from corresponding papers and normalized to U280 while others are proportionally scaled with the number of memory channels.}
\item{*Note the maximal LUT usage in practice is less than 80\%.}
\end{tablenotes}
\end{table}

It turns out that the effective utilization of HBM requires significant logic resources.
Table~\ref{table:resource_projection} presents the estimated resource utilization of existing designs on U280~\cite{xilinxu280}, the largest HBM-enabled FPGA on the market, where we obtain the resource usage from corresponding papers and proportionally project them with the number of utilized memory channels.
{All designs largely exceed the resource capacity even when only 8 of the 32 memory channels are used.
This underutilization of the HBM implies that performance can be scaled further if we find a way to use the logic resources more efficiently.}

The key contribution of our work is in finding the opportunity to be more resource-efficient by taking into account the diversity of workloads in graph processing.
While graph partitioning is a widely employed technique for improving memory efficiency and extracting data-level parallelism~\cite{Zhou2018CF,Zhou2016,ForeGraph,graphgen,zhou2019hitgraph,chen2019fly,chen2021thundergp,isca2021,ruoshi2019improving}, partitions are inevitably unbalanced due to the irregular graph structure~\cite{graphpartitioning,Chaos,sun2017graphgrind}. 
As a result, different partitions can have quite different requirements.
For example, in a pull-based execution model, every vertex reads vertex properties from its neighbors. 
{A partition containing more vertices with high in-degrees tends to have more memory accesses to the vertex property array, hence a better data locality.} 
Previous research used monolithic pipelines that employ elaborate techniques to provide high performance for a wide variety of graph partitions~\cite{chen2021thundergp,isca2021,zhou2019hitgraph}. 
However, this can lead to over-provisioned pipeline designs and underutilization of hardware resources. 
Partitions with poor locality require a different memory access technique that is not necessary for partitions with good locality.
This leads to the key idea of our proposal: heterogeneous pipelines.


\blue{While heterogeneity has been widely adopted in multi-core architectures~\cite{kumar2004single,hma,mittal2016survey}, 
realizing heterogeneous pipelines to accelerate graph processing on FPGA is nontrivial.
First, due to the irregularity of graphs, there exists significant workload diversity within graph processing, which in turn leads to a large design space of pipeline types.
Second, the pipelines must be efficiently tailored to the challenging irregular memory access patterns of graph processing while being resource-efficient.
Third, the computational pattern of graph processing is data-dependent, which makes scheduling tasks for the heterogeneous pipelines even more challenging.}
\blue{In this paper, we  tackle the above-mentioned challenges and propose a heterogeneous pipeline architecture for graph processing that efficiently adapts to workload diversity.}

In particular, we make the following contributions.
\begin{itemize}[leftmargin=*]
    \item We classify graph partitions to dense and sparse partitions by grouping vertices based on their degrees; The dense partitions have high-degree vertices, with good locality, and the sparse partitions have low-degree vertices, with low locality. Then, on the basis of the workload characteristics, we customize two types of pipelines: Little and Big pipelines.
    \item We propose a model-guided task scheduling method that maps partitions to suitable pipeline types based on the proposed performance model, determines the most efficient pipeline combination, and balances the workloads of pipelines.
    \item To ease the entire process, we develop an open-sourced end-to-end framework~--~ReGraph, which generates efficient deployable graph accelerators with user-defined functions and schedules graphs in a push-button manner. 
    \item {The comprehensive evaluation shows ReGraph delivers 1.6$\times$--5.9$\times$ performance speedup and 12$\times$ resource efficiency improvement over state-of-the-arts. \blue{ReGraph is 1.5$\times$--9.7$\times$ faster than the state-of-the-art CPU solution and 2.5$\times$--9.2$\times$  more energy efficient than GPUs.}}
\end{itemize}



\section{Maximizing Performance Per Resource with Heterogeneous Pipelines}\label{sec:motivation}

\blue{By efficiently adapting to the diversity of applications, {\em heterogeneous multi-core architectures} (HMAs) deliver greater throughput for a given silicon area~\cite{kumar2004single,hma,mittal2016survey}. 
While HMA has been effective for multi-core architecture, we need to revisit the challenges and opportunities of heterogeneous design for graph processing on FPGAs. In this section, we study
the workload diversity in graph processing and explore lightweight heterogeneous pipelines with maximized performance per resource. 
In particular, we characterize the diversity of workloads in graph processing, classify the workloads into two different categories, and specialize two types of pipelines to these workload characteristics for higher resource efficiency.}




\subsection{Workload Classification: Dense \textit{vs.} Sparse}\label{sec:graphprocessing}

\begin{figure}[t!]
  \centering
  \includegraphics[width=1\linewidth]{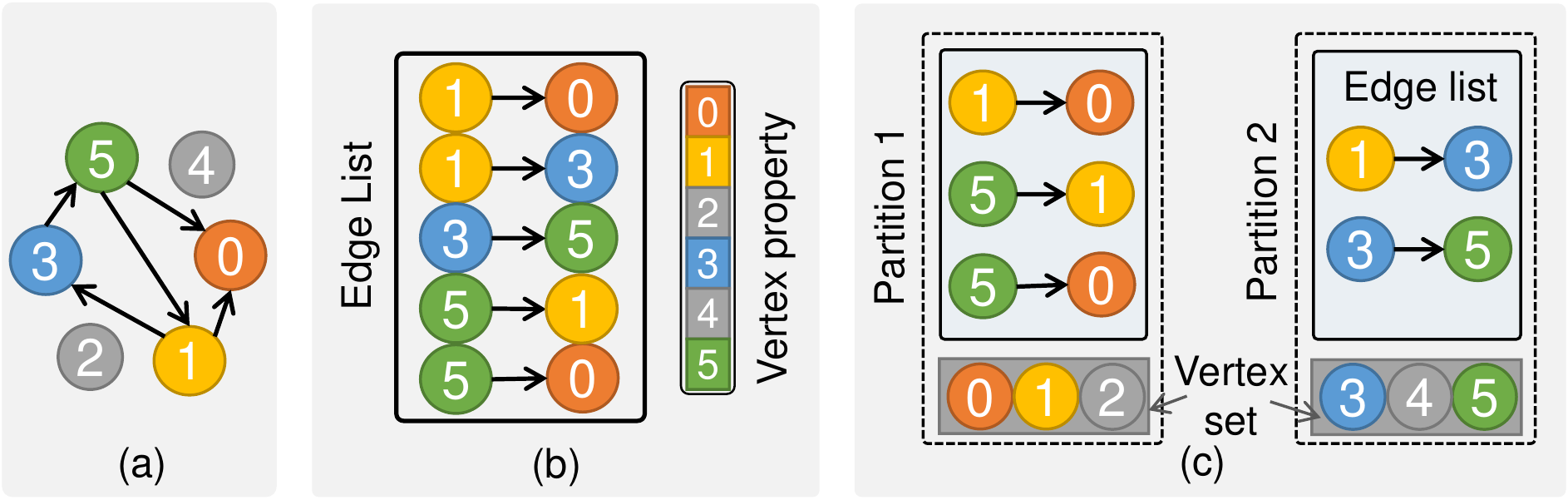}
  \caption{\blue{(a) The example graph; (b) The standard COO graph representation; (c) Graph partitioning on the example graph, assuming the size of vertex set is three.}}
  \label{fig:gasmodel}
\end{figure}

Graphs can be processed in a vertex-centric or edge-centric manner, with the latter being the more popular of the two~\cite{Zhou2018CF,Zhou2016,ForeGraph,zhou2019hitgraph,chen2019fly,chen2021thundergp,isca2021,ruoshi2019improving}. 
In edge-centric processing, edges are accessed sequentially with high memory efficiency.
To access vertices efficiently, the vertices of large graphs are usually partitioned to fit into the limited on-chip RAMs to avoid random memory accesses.
Most state-of-the-art designs~\cite{isca2021,chen2021thundergp} opted to buffer destination vertices and employ customized memory access techniques to access source vertices from the global memory, as buffering both results in a large amount of redundant data transfers.  
Figure~\ref{fig:gasmodel} illustrates the graph partitioning method of ThunderGP~\cite{chen2021thundergp}. 
{The input is a directed graph in standard coordinate list (COO) format with the row indices (source vertices) in ascending order}~\cite{chen2021thundergp,zhou2019hitgraph}. 
Suppose a graph has $V$ vertices ($V$=6 in the example) and the size of vertex set of a partition is $U$ ($U$=3 in the example), $\lceil V/U \rceil$ partitions will be generated with the vertex set of the $i^{th}$ partition ranging from $(i-1) \times U$ to $i \times U$.
Partitions also maintain edge lists that contain all edges whose destination vertices belong to the vertex set. 
In this paper, we adopt this graph partitioning method.

\begin{figure*}[btp]
  \centering
  \includegraphics[width=\linewidth]{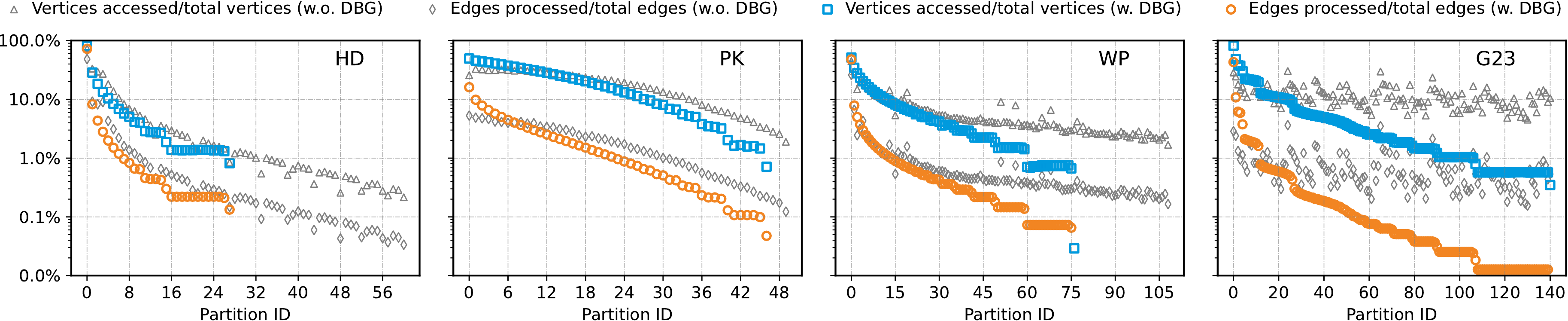}
  \caption{\blue{Workload characteristics of partitions with and without vertex grouping (DBG).}}
  \label{fig:partition_profile}
  \vspace{-6mm}
\end{figure*}

Graph partitions are naturally imbalanced because most graphs are naturally irregular~\cite{graphpartitioning,sun2017graphgrind,beamer2015locality}.
Power-law graphs usually have a few high degree vertices (hot vertices) that are involved in lots of connections~\cite{graphchallenges,DBG}. 
Therefore, the distribution of these vertices influences the workloads of partitions significantly.
Figure~\ref{fig:partition_profile} shows workload characteristics of graph partitions from four representative datasets (see Table~\ref{table1:graphdataset}). 
Note that the y-axis of the figure is on a logarithmic scale.
For each partition, we profile the percentage of edges and the percentage of source vertices accessed. 
The figure suggests large workload diversity in the graph partitions.

{Although graph workloads are diverse, we note that they can be easily divided into two categories, namely one with high-degree vertices and the other with low-degree vertices.
This can be achieved using the lightweight {\em degree-based grouping} (DBG)~\cite{DBG} technique, which is widely used to balance partitions and improve cache efficiency~\cite{isca2021,DBG,chen2021workload}.}
The colored markers in Figure~\ref{fig:partition_profile} depict partition workload characteristics with the partitions sorted in the descending order of in-degree after applying DBG. 
Partitions without any edges are not included. 
We see that we can categorize them into two major kinds of partitions:
1) \textbf{dense partitions} are partitions that have a large number of edges and access a large portion of source vertices.
It happened to be the first few partitions as they contain all high degree vertices. For example, the first partition of HD covers up to 72\% of edges and accesses 80\% of source vertices;
2) \textbf{sparse partitions} are partitions that have a few edges and only access a small portion of source vertices.
The remaining partitions are sparse as they only have low degree vertices. For instance, the majority of G23 partitions have only fewer than 1\% of edges and access less than 10\% of source vertices.
{The exact classification of the partitions is determined during the task scheduling stage according to the performance models of two types of pipelines (see Section~\ref{sec:performancemodelingtaskscheduling}).}


\subsection{Rationales for Heterogeneous Pipeline Customization}
We aim at addressing the scalability issue of HBM-enabled FPGAs. 
The ability to easily classify the diverse workloads motivates us to propose two types of heterogeneous pipelines - one for dense and the other for sparse partitions. 

{\textbf{Big pipelines}} are designed to handle sparse partitions.
Firstly, due to the extremely poor data locality of sparse partitions, memory access techniques such as caching, prefetching~\cite{chen2021thundergp} and the {cache miss optimized memory} system~\cite{isca2021} do not work efficiently. 
Therefore, Big pipelines opt to tolerate the latency of inevitable memory requests instead of equipping with resource-intensive memory access techniques. 
Secondly, because sparse partitions have a few edges but are of a large number, partition switching overhead introduced by emptying the pipeline and enqueuing tasks are non-negligible compared to its short execution time~\cite{chen2021thundergp}. 
This severely diminishes speedup from multiple pipelines. 
To mitigate the overhead, we adopt the data routing technique~\cite{chen2019fly,chen2021thundergp} to let Big pipelines process multiple partitions per execution.

{\textbf{Little pipelines}} are designed to process dense partitions.
On the one hand, benefiting from good spatial locality from a large amount of source vertex accesses, Little pipelines read source vertices in a burst manner without concerning redundant memory accesses argued in existing works~\cite{isca2021,chen2021thundergp}.
Therefore, without using resource-intensive memory access techniques~\cite{chen2021thundergp,isca2021}, Little pipelines only adopt a lightweight ping-pong buffer to overlap the source vertex access and edge process. 
On the other hand, since the number of dense partitions is small, and they have a long execution time, the overhead of partition switching is negligible. 
Therefore, we do not adopt the data routing technique to Little pipelines. 

As our pipelines are designed for heterogeneous workloads, they omit underutilized hardware logic in existing pipelines and therefore provide higher performance with more efficient resource utilization. This enables us to scale the graph processing system on HBM-enabled platforms by instantiating more high-performance pipeline instances.



\begin{figure*}[t]
    \centering
    \begin{overpic}[width=1\textwidth,keepaspectratio]{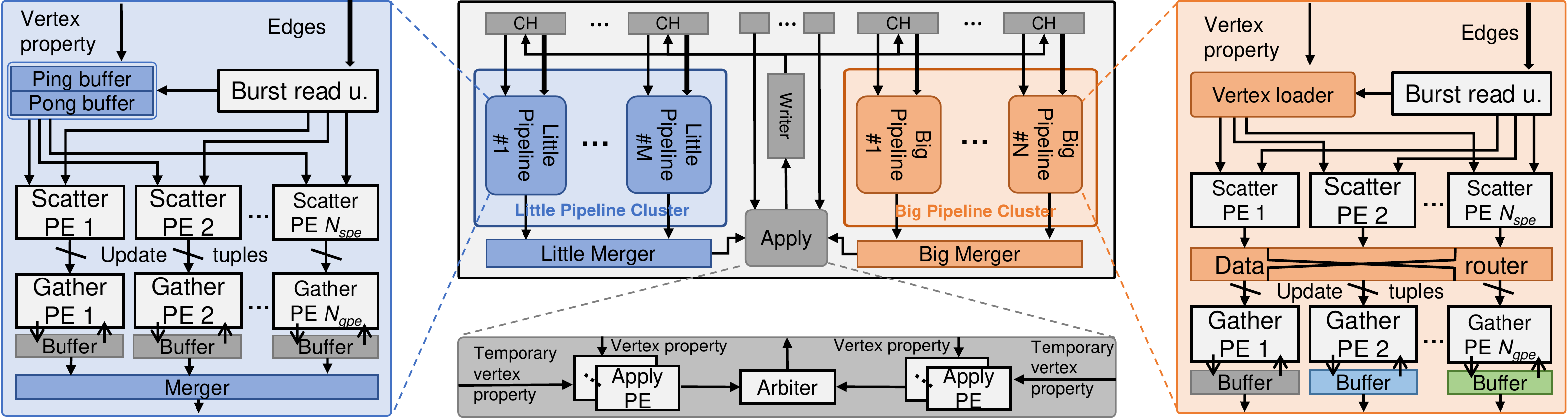}
     \put(42,6.7){(a) Overall architecture}
     \put(1,-2){(b) Little pipeline architecture}
     \put(42,-2){(c) Apply architecture}
     \put(76.5,-2){(d) Big pipeline architecture}
    \end{overpic} 
    \vspace{3mm}
    \caption{\blue{The Big-Little pipeline and Apply architectures, as well as the overview of architecture. Arrows indicate FIFOs.}}
    \label{fig:architectureoverview}
    \vspace{-4mm}
\end{figure*}

\begin{figure}[t!]
  \centering
  \includegraphics[width=1\linewidth]{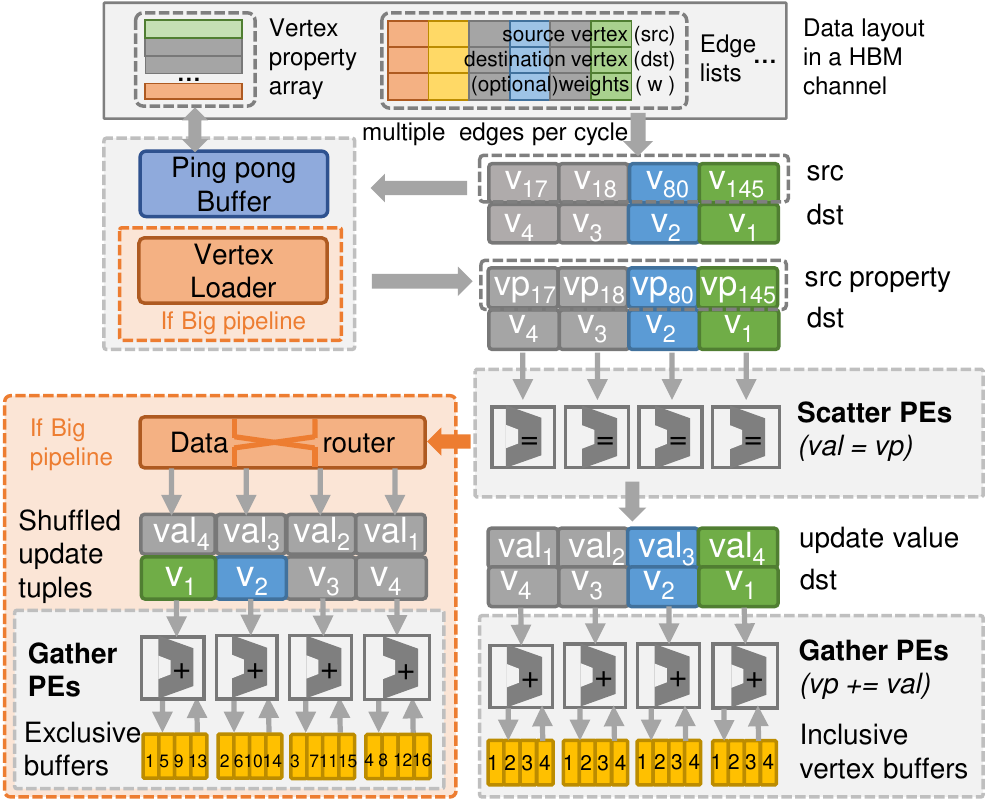}
  \caption{\blue{Data layout in a HBM channel and a running example of PageRank on the Big and Little pipelines. We assume that there are four Scatter PEs and Gather PEs. The output logic for buffered data is not shown.}}
  \label{fig:runingexample}
\end{figure}


\section{System Architecture}\label{sec:systemdesign}
To support various graph algorithms, our system adopts the popular Gather-Apply-Scatter (GAS) model, which contains three stages for each iteration: the \textit{Scatter}, the \textit{Gather}, and the \textit{Apply}~\cite{Zhou2018CF,Zhou2016,ForeGraph,graphgen,zhou2019hitgraph,chen2019fly,chen2021thundergp,isca2021,ruoshi2019improving}.
In this section, we present the proposed heterogeneous pipeline architecture for resource-efficient graph processing.

\subsection{Architecture Overview}~\label{sec:archoverview}
Figure~\ref{fig:architectureoverview}a shows the overview of the proposed architecture. It is composed of a Little pipeline cluster with $M$ Little pipelines, a Big pipeline cluster with $N$ Big pipelines, the Little and Big mergers, the Apply and the Writer modules. 

Little and Big pipelines connect to disjoint memory channels and perform the \textit{Scatter} and the \textit{Gather} stages for dense partitions and sparse partitions, respectively. 
\blue{Both of them manipulate $N_{spe}$ Scatter PEs and $N_{gpe}$ Gather PEs to process multiple edges per cycle and consume the full bandwidth of a memory channel. 
Figure~\ref{fig:runingexample} shows the data layout in a HBM channel and a running example.
The input is a set of edges composed of source vertex ID, destination vertex ID and weights (optional).}
\blue{
The {\textit{Scatter}} stage calculates update values for destination vertices by processing the source vertex properties ({retrieved from the global memory by dereferencing source vertex ID}). 
The {\textit{Gather}} stage accumulates update values for destination vertices whose temporary properties are buffered in {local buffers} and writes out the accumulated values after all edges of the current task are processed. 
The Big and Little mergers combine the intermediate results in buffers of the Big and Little pipelines, respectively. 
}

\blue{
The Apply module receives accumulated temporary results from the Big and Little pipeline clusters simultaneously, as shown in Figure~\ref{fig:architectureoverview}c.
Together with vertex properties from HBM channels, it calculates new vertex properties with multiple PEs.
The new vertex properties are transferred to the Writer on a first-come-first-serve basis.
The Writer finally writes new vertex properties to all memory channels for the next iteration. 
}
All accesses to the global memory are in granularity of a block {(with 512-bit)} for high memory efficiency.


\subsection{Big Pipeline Architecture}~\label{sec:bigpipedesign}
Figure~\ref{fig:architectureoverview}d depicts the architecture of the Big pipeline, which is composed of the Burst read module, the Vertex Loader, the Data Router, $N_{spe}$  Scatter PEs, and $N_{gpe}$  Gather PEs. 
\blue{Figure~\ref{fig:runingexample} shows a running example of PageRank on the pipeline.
The Burst read module sequentially reads multiple edges and duplicates source vertices {for the Vertex Loader}. 
The Vertex Loader retrieves source vertex properties for Scatter PEs by tolerating memory access latency.
The Data Router dynamically dispatches update tuples generated by Scatter PEs to Gather PEs that buffer the corresponding destination vertices.
This allows Gather PEs to process and buffer distinct vertices; therefore, $N_{gpe}$ Gather PEs can handle $N_{gpe}$ partitions per execution (while Little pipelines only handle one), minimizing the number of partition switches. We adopt a multi-stage butterfly network in the Data Router for high resource efficiency~\cite{choi2021hbm,chen2019fly}.}
Next, we introduce details of the Vertex Loader.

\begin{figure}[t!]
  \centering
  \includegraphics[width=0.95\linewidth]{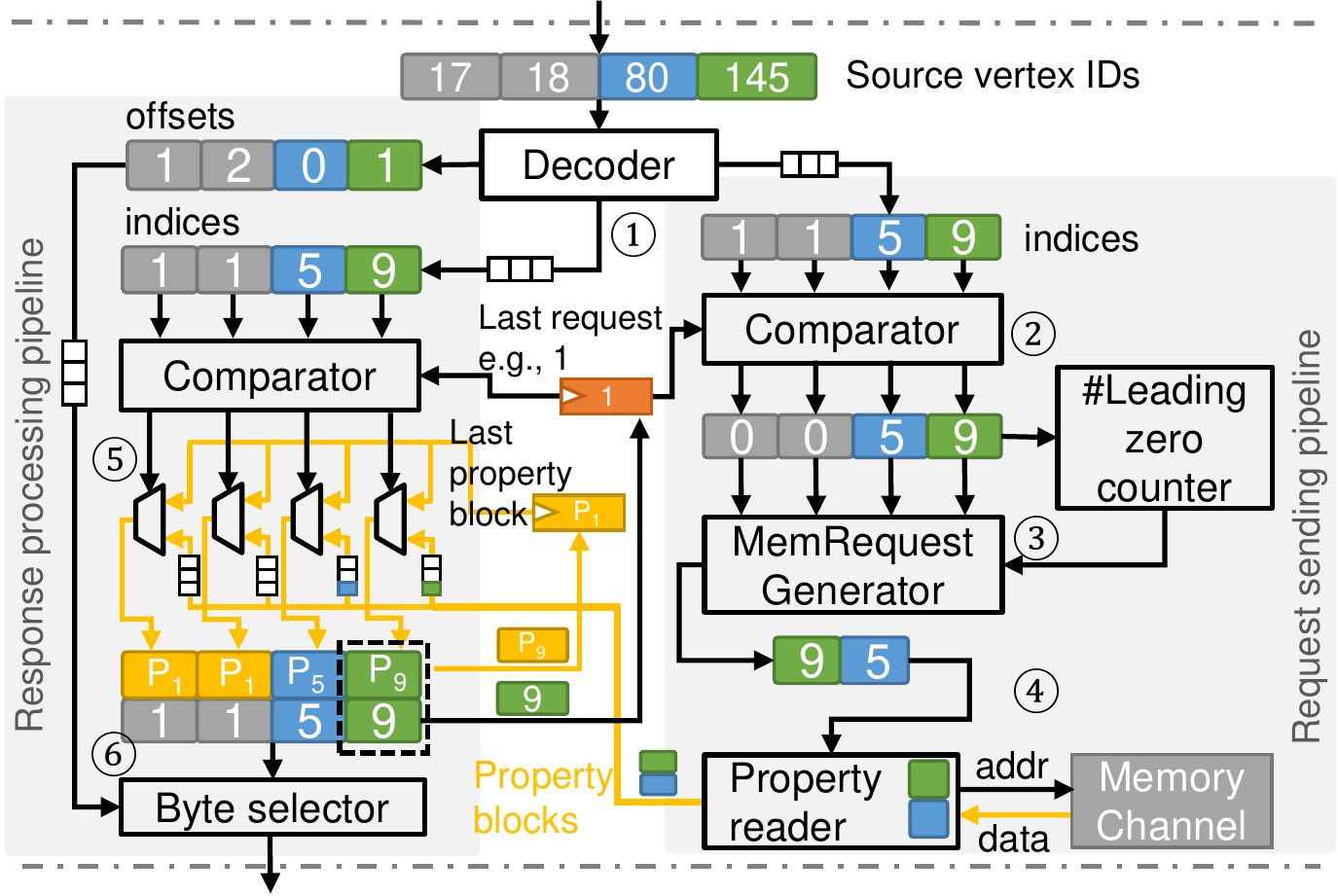}
  \caption{\blue{The architecture and data flow of the Vertex Loader, assuming the Big pipeline processes four edges per cycle.}}
  \label{fig:streambuffer}
\end{figure}

Figure~\ref{fig:streambuffer} shows the architecture and data flow of the Vertex Loader, where we assume there are four Scatter PEs processing four edges per cycle. 
The input is a set of source vertex IDs (four in the example) extracted from a set of edges. 
The output is a set of source vertex properties that Scatter PEs are requesting.
\blue{As the IDs are in ascending order with the standard COO graph formats, we only cache the {last request of the previous vertex ID set} (assumed as one in  Figure~\ref{fig:streambuffer}).}
The logic is split into two small pipelines: the Request sending pipeline that minimizes the number of issued memory requests to the global memory, and the Response processing pipeline that dispatches fetched source vertex properties to Scatter PEs in parallel.
This allows execute/access decoupling as memory requests can be issued before processing. 


\blue{As shown in the Figure~\ref{fig:streambuffer}, the data flow of the Vertex Loader is as follows.
In Step~\ding{172}, the Decoder module calculates block indices in the global memory and the offsets of vertices in the blocks in parallel. 
For example, if the vertex property is 32-bit in the memory, the indices equal to $ \lfloor src \cdot 32/512 \rfloor$ and offsets equal to $ src \cdot 32 \mod 512$.
}
In Step~\ding{173}, the Request sending pipeline compares indices with the last requested index (one in this example) and marks it as zero if matched, otherwise outputs the index.
In Step~\ding{174}, the memory Request generator extracts valid memory requests (non-zeros). As indices are monotonically increased, it ascertains the positions of valid requests by counting the number of leading zeros.
In the example, the index set has two zeros; thus, the Request generator reads the requests from the offset of two to the end. 
As a return, it saves two cycles compared to enumerating all indices.
\blue{
In Step~\ding{175}, the Property reader fetches a vertex property block (512-bit as well) for each block index from the global memory, and writes it to the corresponding stream in a blocking manner based on its offset in the current index set. 
}
In the example, two property data blocks are written to the third stream and the fourth stream, respectively, as their offsets are three and four. 

The Response processing pipeline responses source vertex properties for $N_{spe}$ Scatter PEs in one cycle. 
In Step~\ding{176}, it compares the block indices with the last request in parallel and reuses the last requested property block if they are matched, otherwise reads from the stream. In the example, it only reads the third and fourth streams as the first two indices are matched.  
In Step~\ding{177}, the pipeline decodes out the vertex properties based on their offsets in the corresponding property blocks and sends them to Scatter PEs in parallel.
Lastly, the last request index and its property block are updated with the last index of the current set and its properties, respectively.

\subsection{Little Pipeline Architecture}~\label{sec:littlepipedesign}
Figure~\ref{fig:architectureoverview}a shows the architecture of the Little pipeline, which contains the Burst read module, the Ping-Pong Buffer, the Merger, $N_{spe}$ Scatter PEs, and $N_{gpe}$ Gather PEs.
\blue{Figure~\ref{fig:runingexample} shows a running example of PageRank on the pipeline. 
The Burst read unit sequentially reads edges. 
The Ping-Pong Buffer accesses source vertex properties for Scatter PEs in a burst manner. 
Without dynamic data routing, the update tuples generated by Scatter PEs are statically dispatched to Gather PEs.
As different Gather PEs process update tuples with the same destination vertices, they buffer the same destination vertices, as shown in Figure~\ref{fig:runingexample}.
As a consequence, the Merger accumulates their intermediate results once all edges of a partition are processed. Instead, Big pipelines do not require merger as PEs process distinctive vertices.}
Next, we introduce the detailed HLS-based design of it.

\begin{figure}[t!]
  \centering
  \includegraphics[width=0.9\linewidth]{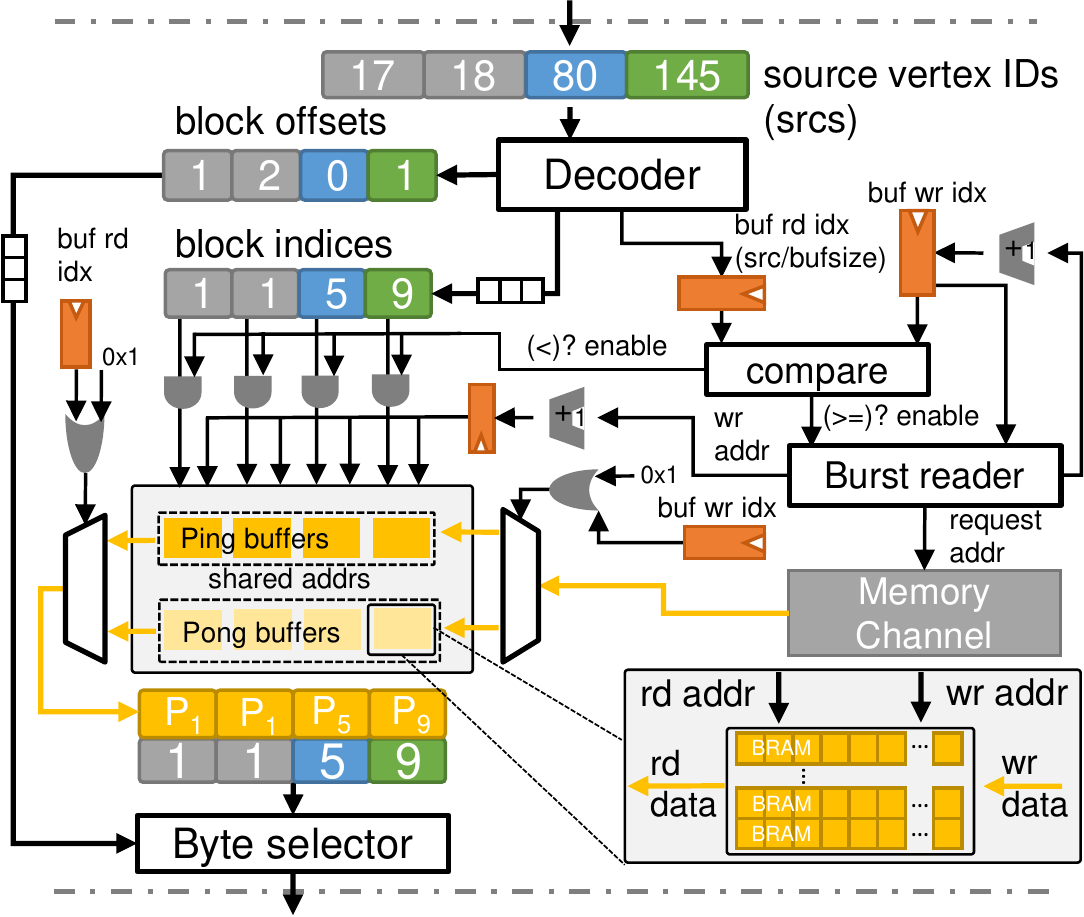}
  \caption{\blue{The architecture and data flow of the Ping-Pong Buffer, assuming the Little pipeline processes four edges per cycle.}}
  \label{fig:ppbarch}
\end{figure}

Ping-pong Buffer allows the pipeline to read vertex properties from one buffer and meanwhile fetch vertex properties from the global memory to the other buffer, hence improving effective memory bandwidth. 
Figure~\ref{fig:ppbarch} shows the proposed Ping-Pong Buffer architecture. 
Same with the Vertex Loader in Big pipelines, inputs are source vertex IDs, and outputs are vertex properties. 
We allocate both ping and pong buffers for each Scatter PE for parallel processing.
\blue{To enable 512-bit data access to a buffer, we cascade multiple BRAMs to construct a 512-bit memory port, e.g., eight BRAMs for Xilinx devices (72 bit $\times$ 8), as shown in the bottom right of Figure~\ref{fig:ppbarch}.}
The logic for filling and reading buffers are realized in one loop with an {\em initiation interval} (II)~\cite{vitis} of one.
The buffer write index and the buffer read index are used for synchronization, and they are initialized as zeros. 
The read index is calculated by dividing the vertex ID by the buffer size. 
The write index is determined by the read index as the buffers are written before read. 
Switching between two buffers is determined by the last bits of the write and read indices.

Buffer filling is executed only when the write index is behind the read index or not ahead by one (to avoid override of the other buffer).
\blue{The Burst reader is responsible for writing successive data blocks to buffers and accessing the global memory in a burst manner.
In each cycle, it requests one data block to buffers.
Once the buffers (e.g., ping buffers) are full, it increases the buffer write index by one. This will switch buffering filling to other buffers in the next execution (e.g., pong buffers).} 
On the other hand, the pipeline reads vertex property blocks for Scatter PEs when the read index is behind the write index, indicating that the properties are loaded into the buffers.
Multiple property blocks can be returned per cycle with duplicated buffers.
The Byte selector outputs the vertex properties based on block offsets in corresponding vertex property blocks.  
As vertex property access addresses are monolithic increased, the architecture also adopts a jump access mechanism that forces the write index to be the read index (not shown in the figure). This avoids redundant memory accesses when the Little pipeline processes only a portion of the partition.

\section{\blue{Graph-Aware Task Scheduling}}\label{sec:performancemodelingtaskscheduling}

While the effectiveness of heterogeneous architectures heavily depends on task scheduling, an accurate performance model that estimates the execution time of partitions on both types of pipelines is required for effective partition-to-pipeline mapping (i.e., identifying whether a given partition is dense or sparse) and workload balancing. 
In this section, we first introduce the performance model of two types of pipelines and then the model-guided task scheduling method.

\subsection{Performance Modeling of Big-Little Pipelines}\label{sec:performancemodeling}
Unlike regular applications that have deterministic memory access and computation latency~\cite{cong2018automated}, irregular graph structure makes performance modeling challenging. 
A simple regression model based on the numbers of edge and vertex is unable to model the processing performance accurately~\cite{chen2021thundergp}, because the bottleneck of the pipeline alternates between edge access and vertex access during execution. 
\blue{Instead, we propose a cycle-level performance model that accurately estimates the execution time of Big and Little pipelines of an application on graph partitions by enumerating edges.
As performance estimation has only lightweight computation and is integrated to the graph partitioning phase to reduce edge enumeration overhead, it introduces little extra preprocessing overhead.} 

The estimated execution cycles $C_p$ of two types of pipelines on a partition $p$ is shown in Equation~(\ref{eq:1b}):
\begin{equation}\label{eq:1b}
C_p = \sum_{i=0}^{E_p}{\max (C_{acs\_v}^i, C_{acs\_e}, {C_{proc}})} + C_{store} + C_{const}
\end{equation}
where $i$ enumerates $E_p$ edges of a partition (happening with graph partitioning), $C_{acs\_v}^i$ denotes cycles to access the source vertex of the edge, $C_{acs\_e}$ denotes cycles of reading the edge, {$C_{proc}$} represents cycles to process the data, $C_{store}$ denotes cycles to write out buffered destination vertices and $C_{const}$ is the constant overhead.

As edges are sequentially accessed, given the data size the memory channel can access in one cycle (i.e., the size of the data block), $S_{mem}$, and the size of an edge, ${S_e}$, $C_{acs\_e}$ can be calculated as a constant value, $\frac{S_{e}}{S_{mem}}$.
For $C_{const}$, we measure the execution time of dummy partitions with a few edges to estimate the constant overhead of partition switching.  


Let $S_{ram}$ denote the {data width of the port} of the buffers in $N_{gpe}$ Gather PEs and let $S_{buf}$ denote the size of the buffer. 
$C_{store}$ is calculated by Equation~(\ref{eq:store}). 
The buffers of Gather PEs in Little pipelines are merged; hence, the data size to write out is $N_{gpe}$ times smaller than that of the Big pipeline. 
\begin{equation}\label{eq:store}
C_{store} = 
\begin{cases}
\max(\frac{S_{buf}}{S_{ram}}, \frac{S_{ram}\cdot N_{gpe}}{S_{mem}}), & \text{if Big pipeline} \\
\max(\frac{S_{buf}}{S_{ram}}, \frac{S_{ram}}{S_{mem}}), & \text{if Little pipeline} 
\end{cases}
\end{equation}

Meanwhile, $C_{proc}$ is determined by numbers of Scatter PEs ($N_{spe}$), Gather PEs ($N_{gpe}$) and their IIs ($II_{spe}$ and  $II_{gpe}$), as shown in Equation~(\ref{eq:cproc}). 
The II indicates the number of cycles the PE could process one input and is determined by the compiler once the logic of PE is set.
\begin{equation}\label{eq:cproc}
\frac{1}{C_{proc}} = \max (\frac{N_{spe}}{II_{spe}}, \frac{N_{gpe}}{II_{gpe}})
\end{equation}

Lastly, we model $C_{acs\_v}^i$ based on the architecture of the Vertex Loader and Ping-Pong Buffer in Big and Little pipelines, respectively. 
As the Vertex Loader directly accesses the memory for different requests without caching and prefetching, we benchmark the memory access latency with varying access distance (stride) on the test FPGAs~\cite{huang2021shuhai}.
{The benchmark results show that the $C_{acs\_v}^i$ of the Big pipeline can be modeled by a linear function with respect to access distance, as shown in Equation~(\ref{eq:3}), where $a$ and $b$ denote the coefficients, $S_{vprop}$ denote the size of the vertex property and $vid$ the source vertex ID of the edge.}  
In addition, it has an upper bound and a lower bound, as there exists the worst-case and best-case memory access latency.
For the Little pipeline, the Ping-Pong Buffer sequentially reads the vertices; hence, the $C_{src}^i$ of the Little pipeline can be modeled by the access distance and the data size the memory channel can read per cycle, ${S_{mem}}$, as shown in Equation~(\ref{eq:3}). 
\begin{equation}\label{eq:3}
\begin{aligned}
C_{acs\_v}^i = &
\begin{cases}
a \cdot ({vid}^i - {vid}^{i-1}) \cdot S_{vprop} + b, &\text{if Big pipeline} \\
\frac{({vid}^i - {vid}^{i-1}) \cdot S_{vprop}}{S_{mem}}, & \text{if Little pipeline}
\end{cases}
\end{aligned}
\end{equation}

Combining Equations~(\ref{eq:1b})--(\ref{eq:3}), we can estimate the execution cycles of Big and Little pipelines for a given partition. 

\subsection{Model-Guided Task Scheduling}\label{sec:partscheduling}
\blue{Our task scheduling method includes inter- and intra- cluster task schedulings which are based on the estimated execution time of partitions (obtained by the performance model during the graph partitioning phase) to fully utilize the heterogeneous pipeline architecture for a graph.} 
Firstly, the inter-cluster task scheduling method schedules partitions to the suitable type of pipelines and selects the most efficient pipeline combination to minimize the worst execution time of two clusters. 
Secondly, the intra-cluster task scheduling method cuts partitions to sub-partitions with equal execution times to utilize multiple pipelines within clusters.  
The task scheduling process runs offline and only once to generate a static scheduling plan for a graph on an application.



\begin{figure}[t!]
  \centering
     \begin{overpic}[width=\linewidth,keepaspectratio]{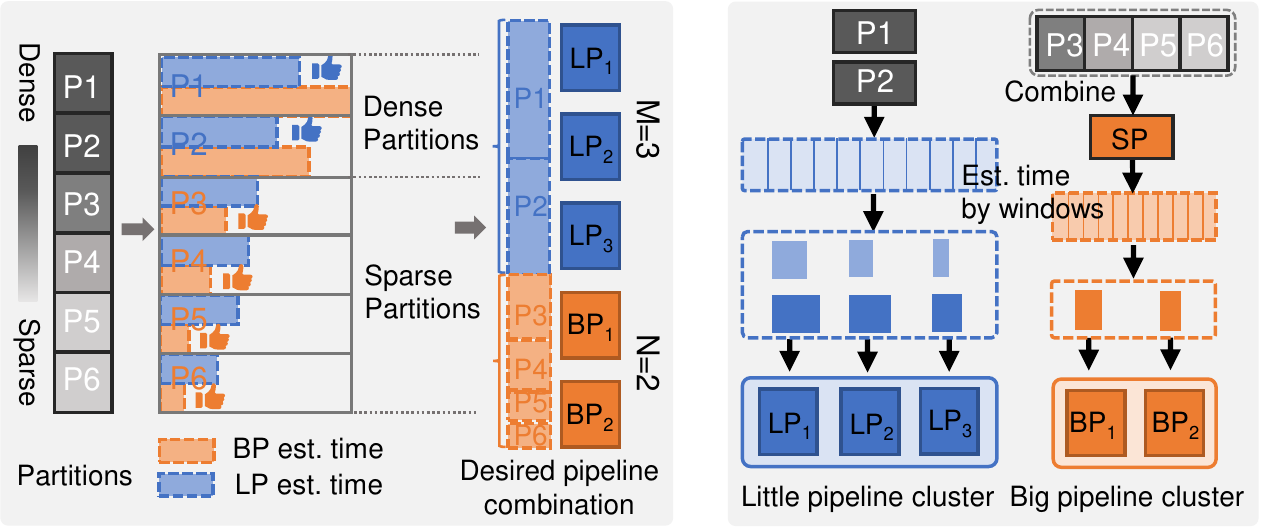}
     \put(5,-4){(a) Inter-cluster scheduling.}
     \put(55,-4){(b) Intra-cluster scheduling.}
    \end{overpic} 
    \vspace{1mm}
  \caption{Model-guided workload balancing. Assuming six partitions, total five pipelines, and $N_{gpe} = 4$.}
  \label{fig:scheduling}
\end{figure}

\begin{figure*}[t!]
  \centering
  \includegraphics[width=1\linewidth]{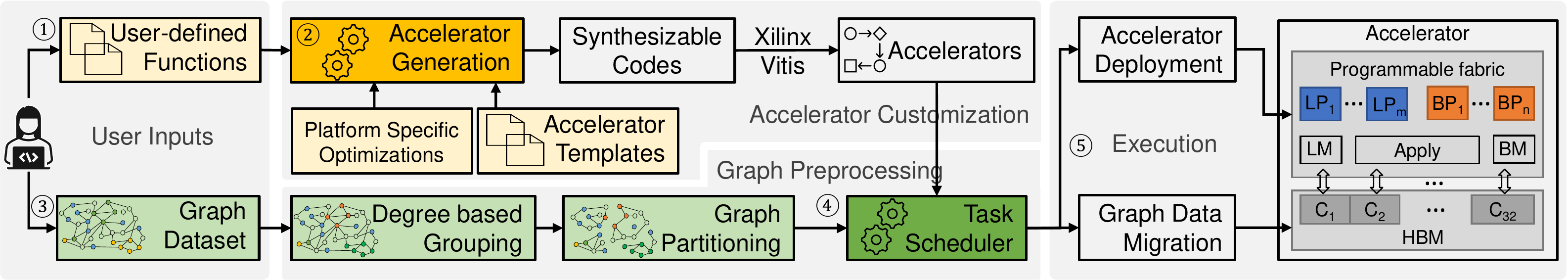}
  \caption{Overview of the workflow with ReGraph.}
  \label{fig:ReGraphoverview}
  \vspace{-5mm}
\end{figure*}

\noindent\textbf{{Inter-cluster task scheduling.}}
Figure~\ref{fig:scheduling}a shows two steps of inter-cluster task scheduling.
Firstly, given a graph with total of $N_{p}$ partitions, it ascertains the number of dense partitions, $X$, and the number of sparse partitions, $Y$, to minimize the overall execution time of partitions on Big and Little pipeline clusters, $\sum_{p=0}^{X}{T_{Little}^p} + \sum_{p=0}^{Y}{T_{Big}^p}$.
{In particular, a partition is marked as a sparse partition if the estimated execution time on the Big pipeline is shorter than that on the Little pipeline, otherwise marked as a dense partition.}
In Figure~\ref{fig:scheduling}, $P_1$ and $P_2$ are marked as dense partitions.

Secondly, it determines the numbers of two types of pipelines to balance the execution time of the Big and Little pipeline clusters.
Assume the total number of pipelines is $N_{pip}$ (bounded by the numbers of memory channels and memory ports of the platform), it sets $M + N = N_{pip}$ and tunes $M$ and $N$ to minimize the difference between execution times of two clusters, $ \big | \frac{\sum_{p=0}^{X}{T_{Little}^p}}{M}-\frac{\sum_{p=0}^{Y}{T_{Big}^p}}{N} \big |$.
Figure~\ref{fig:scheduling}a illustrates the example with a total of five pipelines. Three Little pipelines are allocated to process two dense partitions, whereas two Big pipelines are built for the execution of four sparse partitions. 

\vspace{0.5mm}
\noindent\textbf{{Intra-cluster task scheduling.}}
In our design, pipelines within clusters process a partition cooperatively. 
This requires a partition divided to sub-partitions for multiple pipelines. 
While previous works~\cite{chen2021thundergp,isca2021} cut the edges or vertices of partitions evenly, the irregularity of partitions results in unbalanced execution time of pipelines. 
Instead, we cut partitions to sub-partitions with similar execution times via the proposed performance model.
Figure~\ref{fig:scheduling}b shows the example to cut four sparse partitions for two Big pipelines and two dense partitions for three Little pipelines. 
As the Big pipeline buffers $N_{gpe}$ times as many vertices as the Little pipeline, we merge every $N_{gpe}$ sparse partitions into large sparse partitions before the performance estimation. 
To calculate the boundaries of sub-partitions by scanning once, we estimate execution time at granularity of {a window that contains a certain number of edges} during graph partitioning and then divide these windows into $M$ or $N$ clusters with similar overall execution times.

\section{ReGraph: An Automated Framework}\label{sec:framework}

\subsection{ReGraph Framework Overview}
Figure~\ref{fig:ReGraphoverview} shows the overview of the ReGraph workflow.
To obtain a customized accelerator design for a graph application, developers only need to write {\em user-defined functions} (UDFs) of three stages of the GAS model with the provided programming interface (Step~\ding{172}).
ReGraph then takes the UDFs, accelerator templates and platform specific optimizations to generate a set of synthesizable codes for accelerators with all possible pipeline combinations (Step~\ding{173}). 
The synthesizable codes are compiled to bitstreams using the Xilinx Vitis tool-chain. 


After that, users assign the graph for acceleration (Step~\ding{174}).
ReGraph groups vertices based on their in-degrees and partitions the graph.
Then, the task scheduler with the built-in model-guided task scheduling method selects the accelerator with the most efficient pipeline combination and generates the scheduling plan (Step~\ding{175}).
Lastly, ReGraph deploys the selected accelerator and runs on the target FPGA (Step~\ding{176}).

\subsection{Programming Interface}\label{sec:api}
\begin{lstlisting}[language=C++, caption=User-defined functions for PageRank with ReGraph.,label={listing:codeexample}]
/* logic for the Scatter phase */
inline prop_t accScatter(prop_t srcProp, prop_t edgeProp){
	return (srcProp); }
/* logic for the Gather phase */
inline prop_t accGather(prop_t buf_prop, prop_t value){
	return ((buf_prop) + (value)); }
/* logic for the Apply phase */
inline prop_t accApply( prop_t tProp, prop_t oProp, prop_t outDeg){
	return (((kDampFixPoint * tProp) >> 7) * (1 << 16 ) / outDeg) >> 16; 
/*exit condition is omitted for simplicity */}
\end{lstlisting}

Users can implement different graph accelerators by only writing three high-level functions: {accScatter()}, {accGather()} and {accApply()}.  
We demonstrate users' efforts using PageRank as an example in List~\ref{listing:codeexample}.
In lines 2--5, the {accScatter()} returns the source vertex property, which means the vertex pushes its property (an averaged score) to its neighbours. 
In lines 5--6, the {accGather()} accumulates the property for destination vertices by adding the buffered property and incoming values.
In lines 8--9, the {accApply()} calculates the new property of each vertex by dividing weighted accumulated score by its out-degree.
\blue{For invoking the graph accelerator, ReGraph leverages the OpenCL APIs (enabled by Xilinx Vitis~\cite{vitis}) and provides several encapsulated APIs, e.g., initAccelerator() for initializing hardware context in one function.}


\subsection{Platform-Specific Optimizations}\label{sec:optimizations}
ReGraph adopts various platform-specific optimizations to provide great performance on Xilinx HBM-enabled devices.

\noindent
\textbf{Memory port management.}
Current HBM-enabled FPGA platforms support limited memory ports, e.g., 32 ports on U280, which largely constrains the number of pipelines instantiated on the platform. As a read port and a write port \textit{in one kernel} can be bundled~\cite{vitis}, we propose HBM port wrappers to bundle the write port in the Apply module and the read port for reading vertex properties. 
Wrappers receive memory requests, access the global memory, and send the responses to corresponding modules. 
This optimization reduces memory ports per pipeline from three to two.


\noindent
\textbf{SLR crossing-aware optimizations.} 
Modern FPGAs have multiple \textit{super logic regions} (SLRs) to enlarge resource capacity; however, the costly inter-SLR communication may result in low implementation frequency~\cite{guo2021autobridge,guo2020analysis}.   
Beyond applying the existing timing optimizations~\cite{guo2021autobridge,isca2021}, we implement the Big merger and the Little merger by a merge-tree with many small free running kernels~\cite{vitis} and merge the data within the SLR as much as possible before sending it to other SLRs. 

\noindent
\textbf{Utilizing URAMs.} 
We utilize URAMs for vertex buffering in Gather PEs, using a 64-bit data access granularity. We also solve the read after write hazard by utilizing a set of shift registers to obtain an II of one for Gather PEs.


\subsection{Accelerator Generation}
ReGraph generates a set of accelerators that have different numbers of Big and Little pipelines with the following steps.
Firstly, it tunes the numbers of Scatter and Gather PEs to fully utilize the memory bandwidth of a memory channel.
{Secondly, ReGraph calculates the total number of pipelines that can be instantiated on the platform, $N_{pip}$. 
While resources allow $N_{pip}$ to be the number of memory channels, $N_{ch}$, the number of memory ports, $N_{port}$}, constrains it, as each pipeline occupies two memory ports.
Assume the number of reserved memory ports for the Apply module is $N_{res}$, ReGraph sets $N_{pip}$ as $min(N_{ch}, \frac{N_{port} - N_{res}}{2})$.
Thirdly, ReGraph enumerates the numbers of Big and Little pipelines, by varying $M$ from $0$ to $N_{pip}$ and varying $N$ from $N_{pip}$ to $0$ to generate $N_{pip}$ sets of configurations. 
\blue{Finally, with these configurations, ReGraph spreads the kernels even to SLRs according to a preset kernel-to-SLR mapping table and connects kernels with AXI streams or memory channels.} 
We have developed a python-based code generation program that automatically generates the connectivity of the kernels for synthesizable codes.

\begin{table}[h!]
\centering
\caption{{Two HBM-enabled platforms used in experiments.}}
\label{table:hardwareplatforms}
\resizebox{.48\textwidth}{!}{%
\begin{tabular}{@{}llclllccc}
\toprule
\multicolumn{1}{l|}{Platform}  & \#LUTs  & \#URAMs & \#SLRs    & Bandwidth  & \#CH &  \#Port    & TDP   \\ \midrule
\multicolumn{1}{l|}{Alveo U280 \textbf{(U280)}}  & \multicolumn{1}{c}{1,304K} & \multicolumn{1}{c}{960}   & \multicolumn{1}{c}{3} & \multicolumn{1}{l}{460 GB/s}  & \multicolumn{1}{c}{32} & \multicolumn{1}{c}{32} & \multicolumn{1}{c}{225 W}  \\ \midrule
\multicolumn{1}{l|}{Alveo U50 \textbf{(U50)}}  & \multicolumn{1}{c}{872K} & \multicolumn{1}{c}{640} & \multicolumn{1}{c}{2} & \multicolumn{1}{l}{316 GB/s}  & \multicolumn{1}{c}{32} & \multicolumn{1}{c}{28} & \multicolumn{1}{c}{70 W} \\ \bottomrule
\end{tabular}%
}
\end{table}

\begin{table}[h!]
\centering
\caption{The graph datasets.}
\label{table1:graphdataset}
\resizebox{0.5\textwidth}{!}{%
\begin{tabular}{lrrrrr}
\toprule
\multicolumn{1}{l}{Graphs}  & \text{$\left| V \right|$} &  \text{$\left| E \right|$}&\text{$\left| D \right|$} & Type & Categories \\ 
\midrule

\multicolumn{1}{l|}{ rmat-19-32 (R19)  ~\cite{kronecker2010}} & 524.3K & 16.8M & 32 & Directed & Synthetic  \\
\multicolumn{1}{l|}{ rmat-21-32 (R21)  ~\cite{kronecker2010}} & 2.1M & 67.1M & 32 & Directed & Synthetic \\
\multicolumn{1}{l|}{ rmat-24-16 (R24)  ~\cite{kronecker2010}} & 16.8M & 268.4M & 16 & Directed & Synthetic \\
\multicolumn{1}{l|}{ graph500-scale23 (G23)  ~\cite{graphdataset}} & 4.6M & 258.5M & 56 & Directed & Synthetic  \\
\multicolumn{1}{l|}{ web-google (GG)  ~\cite{graphdataset}} & 916.4K & 5.1M & 6 & Directed & Web \\
\multicolumn{1}{l|}{ amazon-2008 (AM)  ~\cite{graphdataset}} & 735.3K & 5.2M & 7  & Directed & Social \\
\multicolumn{1}{l|}{ web-hudong (HD)  ~\cite{graphdataset}} & 2.0M & 14.9M & 7 & Directed & Web \\
\multicolumn{1}{l|}{ web-baidu-baike (BB)  ~\cite{graphdataset}} & 2.1M & 17.8M & 8 & Directed & Web \\
\multicolumn{1}{l|}{ wiki-topcats (TC)  ~\cite{snapnets}} & 1.8M & 28.5M & 16 & Directed & Web \\
\multicolumn{1}{l|}{ pokec-relationships (PK)  ~\cite{snapnets}} & 1.6M & 30.6M & 19 & Directed & Social \\
\multicolumn{1}{l|}{ soc-flickr-und (FU)  ~\cite{graphdataset}} & 1.7M & \blue{15.6M} & 9 & Undirected & Social  \\
\multicolumn{1}{l|}{ wikipedia-20070206 (WP)  ~\cite{davis2011university}} & 3.6M & 45.0M & 13 & Directed & Web \\
\multicolumn{1}{l|}{ liveJournal (LJ)  ~\cite{snapnets}} & 4.8M & 68.9M & 14 & Undirected & Social \\
\multicolumn{1}{l|}{ ca-hollywood-2009 (HW)  ~\cite{graphdataset}} & 1.1M & \blue{56.3M} & 53 & Undirected & Collabo. \\
\multicolumn{1}{l|}{ dbpedia-link (DB)  ~\cite{graphdataset}} & 18.3M & 172.2M & 9 & Directed & Social \\
\multicolumn{1}{l|}{ orkut (OR)  ~\cite{graphdataset}} & 3.1M & \blue{117.2M} & 38  & Undirected & Social \\
\bottomrule
\end{tabular}
}
\end{table}

\begin{figure*}[t!]
  \centering
  \includegraphics[width=1\linewidth]{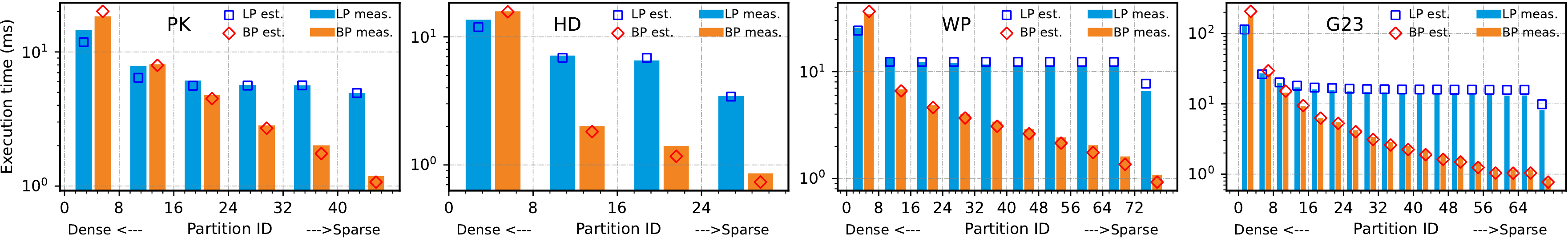}
  \caption{Big and Little pipelines' measured and estimated execution time of PR on partitions of four graphs on U280.}
  \label{fig:model_validation}
\end{figure*}

\section{Evaluation}\label{sec:evaluation}
\blue{We first assess the efficiency of Big and Little pipelines and their performance models. 
We then evaluate the benefits of heterogeneity and {resource utilization}, followed by the demonstration of system scalability and the assessment of the cost of preprocessing.
Finally, we compare ReGraph to state-of-the-art FPGA solutions and CPU/GPU solutions.}

\subsection{Experimental Setup}\label{sec:experimentalsetup}
\noindent
\textbf{Hardware platform.}
Table~\ref{table:hardwareplatforms} shows two HBM-enabled FPGAs used in our evaluation. 
U50 is a low-profile card with fewer resources and a lower {\em thermal design power} (TDP). 
It supports only 28 memory ports, resulting in a lower peak memory bandwidth. 
We host U280 and U50 on servers with Xeon Gold 6246R CPU and Xeon W-2155 CPU, respectively. 
Xilinx Vitis 2020.2 is used for development for both platforms. 

\noindent
\textbf{Applications and datasets.} 
{
We consider three graph processing applications as benchmarks: 
PageRank \textbf{(PR)}, Breadth-First Search \textbf{(BFS)}, 
and Closeness Centrality \textbf{(CC)}.}
{
Table~\ref{table1:graphdataset} shows the details of the used graph datasets, including synthetic~\cite{kronecker2010} graphs and real-world large-scale graphs. 
}

\noindent
\textbf{Implementation details.}
{The prototype of ReGraph consists of 2,787 lines of HLS code for code templates, 2,063 lines of C++ code for graph preprocessing, scheduling and accelerator deployment and 423 lines of Python code for automated accelerator code generation.}

\noindent
\textbf{Parameter details.}
The size of the Ping-Pong Buffer is 32KB.
The depths of streams that cross SLRs are set to 16 for better timing.
For all applications, the numbers of Scatter PEs and Gather PEs (with II of one) of a pipeline are set to eight.
While the resources of the two platforms allow us to instantiate one pipeline per memory channel, the memory port limitation constrains the number of pipelines to 14 on U280 and 12 on U50. 
Each Gather PE buffers 65,536 destination vertices on U280 and  32,768 on U50. 
{All raw graph data are 32-bit in our experiments. 
Same with ThunderGP~\cite{chen2021thundergp} and GraphLily~\cite{hu2021graphlily}, ReGraph uses fixed-point data type for PR.} 

\noindent
\textbf{Baselines.}
ThunderGP~\cite{chen2021thundergp} and Asiatici et al.~\cite{isca2021} are two state-of-the-art graph processing frameworks using multiple SLRs and memory channels on DRAM-FPGA platforms. 
GraphLily~\cite{hu2021graphlily} is a graph linear algebra overlay on HBM-equipped FPGAs that expresses different graph algorithms with two built-in primitives.


\subsection{Efficiency of Big-Little Pipelines and Their Models}\label{sec:modelvalidation}

We first evaluate the performance of two types of pipelines on different partitions together with the proposed performance model.
Figure~\ref{fig:model_validation} presents the measured and estimated execution time of PR of a single Big/Little pipeline on partitions of four graphs (profiled in Figure~\ref{fig:partition_profile}). 
We report execution time per eight partitions, as the Big pipeline processes eight partitions per execution, benefiting from data routing. 

As shown in Figure~\ref{fig:model_validation},
the Little pipeline executes faster than the Big pipeline when the partition is dense (the first few partitions) while the Big pipeline performs better when the partition is sparse (the rest of partitions). 
This is attributed to the architectures of Big and Little pipelines. 
At the same time, the estimated performance is very close to the measured performance. 
The average error ratio (defined as the difference between estimated and measured execution time dividing measured execution time) of the Big pipeline's model is only 4\% and that of the Little pipeline's model is 6\%.

\begin{figure*}[h!]
  \centering
  \includegraphics[width=1\linewidth]{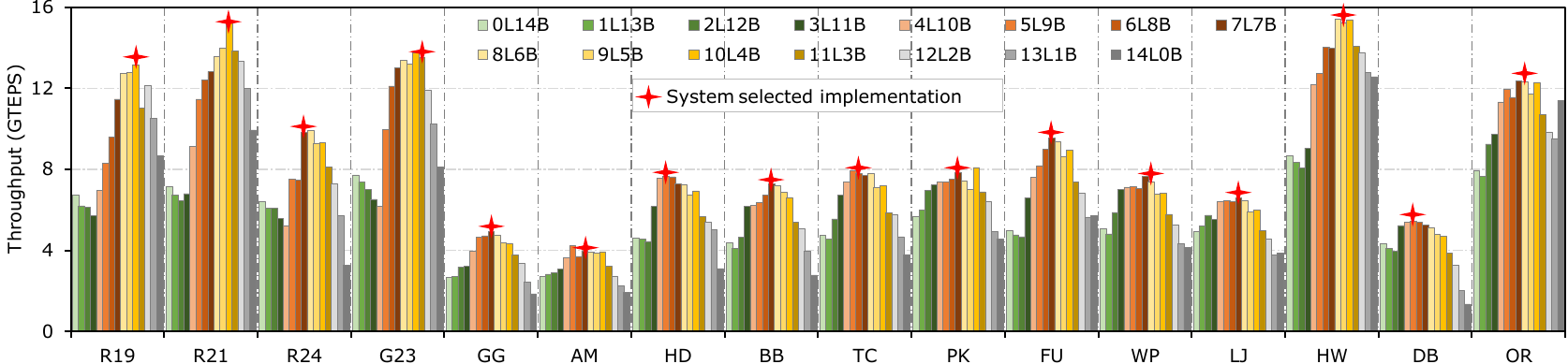}
  \caption{\blue{Performance of PR with varying the numbers of Big (B) and Little (L) pipelines on U280.}}
  \label{fig:dse}
\end{figure*}

\begin{figure*}[btp]
  \vspace{-2mm}
  \centering
  \begin{minipage}[t]{.49\textwidth}
  \vspace{0pt}
  \includegraphics[width=1\linewidth]{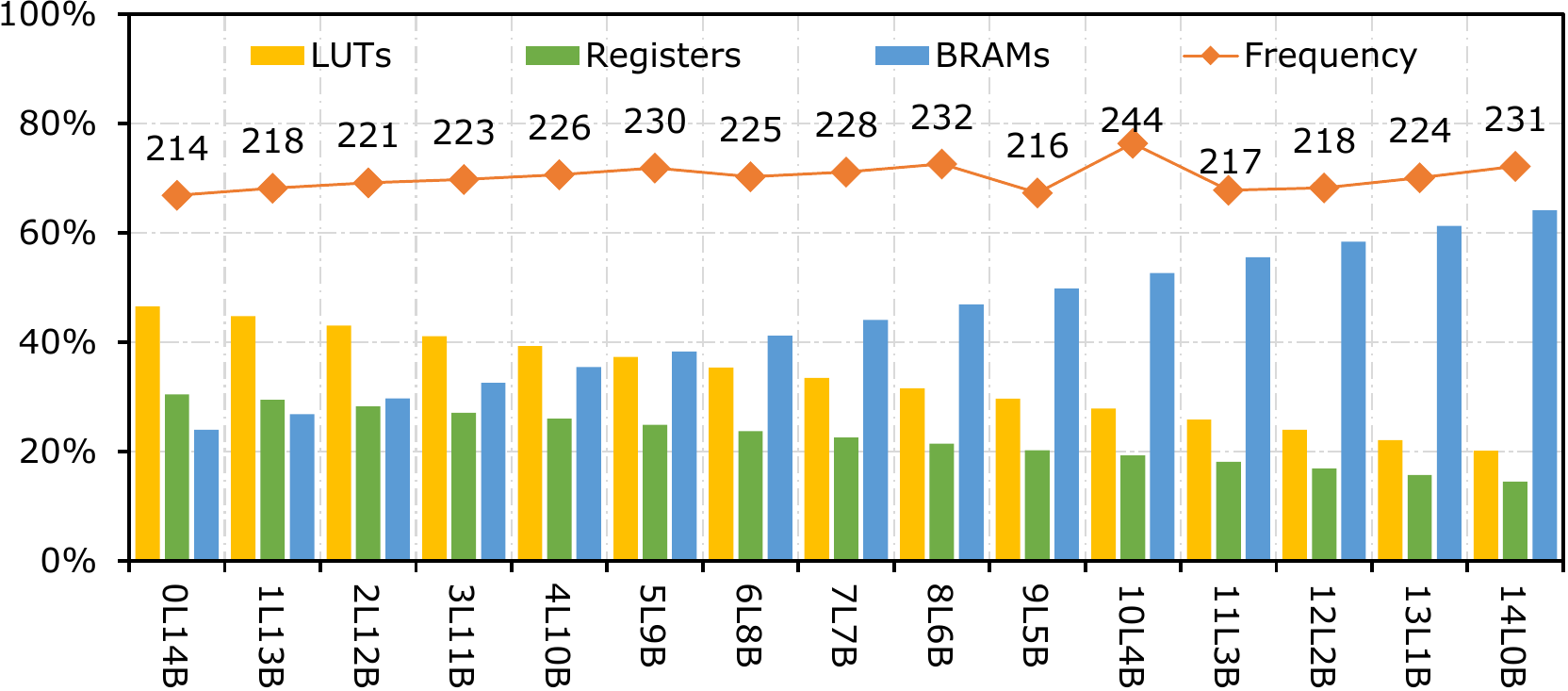}
  \caption{Resource utilization and frequency of PR implementations with all pipeline combinations on U280.}
  \label{fig:pr_resource_u280}
  \end{minipage}
  \hfill
  \begin{minipage}[t]{.5\textwidth}
  \vspace{0pt}
  \includegraphics[width=1\linewidth]{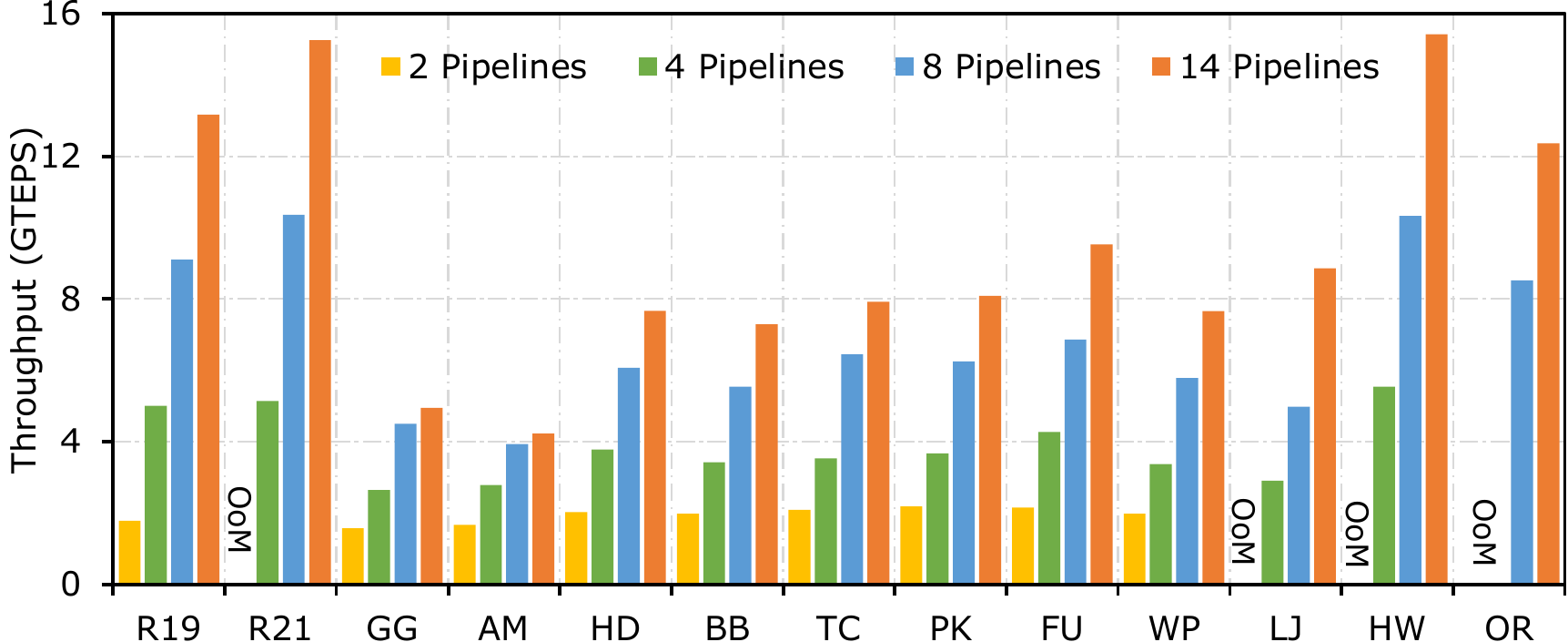}
  \caption{\blue{Performance of PR with varying the number of pipelines on U280.}}
  \label{fig:scalibility}
  \end{minipage}
  \vspace{-2mm}
\end{figure*}

\begin{table*}[t!]
\centering
\caption{\blue{Preprocessing time with one CPU (Xeon Gold 6248R) thread in millisecond.}}
\label{table:preprocessingoverhead}
\resizebox{\textwidth}{!}{%
\begin{tabular}{@{}l|llllllllllllllll@{}}
\toprule
Graphs                  & R19   & R21   & R24    & G23    & GG   & AM   & HD    & BB    & TC    & PK    & FU    & WP    & LJ    & HW     & DB     & OR     \\ \midrule
Vertex Grouping (DBG)   & 3.4   & 14.2  & 111.2  & 29.9   & 9.6  & 7.3  & 12.6  & 18.8  & 13.9  & 14.9  & 10.8  & 28.9  & 34.3  & 7.3    & 131.0    & 30.9   \\ \midrule
Partitioning \& Scheduling & 168.9 & 719.6 & 4054.1 & 2943.3 & 66.1 & 57.0 & 171.1 & 229.4 & 357.1 & 318.9 & 436.5 & 508.9 & 996.3 & 1290.4 & 2842.9 & 2977.1 \\ \bottomrule
\end{tabular}
}
\vspace{-2mm}
\end{table*}

\subsection{\blue{Benefits of Heterogeneity}}\label{sec:benefitsof}
\blue{
Figure~\ref{fig:dse} shows PR implementations with different pipeline combinations, where we vary the numbers of Little and Big pipelines.
The numbers of dense and sparse partitions are determined by the framework. Implementations with only Big pipelines (0L14B) or only Little pipelines (14L0B) are referred as homogeneous pipeline architectures.
}

\blue{
There are several highlights. 
First, the best performance implementations are always with mixed pipeline types rather than a single type, which demonstrate the benefits of heterogeneity. 
Moreover, the configuration on the numbers of Little and Big pipelines varies across different graphs, which demonstrates both Little and Big pipelines contribute to the overall performance.
Second, the performance of our system selected implementations is close (around 92\% on average) to that of the best implementations.
Third, synthetic graphs (R19, R21, R24 and G23) have better performance and require more Little pipelines than real-world graphs, because they are relatively regular and have larger portion of edges located in dense partitions.
}

\subsection{Resource Utilization}\label{sec:resourceutilization}
Figure~\ref{fig:pr_resource_u280} presents resource utilization and frequency of PR with different pipeline combinations on U280.  
We observe similar resource utilization for other applications.
We omit the presentation of URAM utilization as it decides the partition size and is constantly 96\% for all implementations. 
Overall, the highest performant implementations such as 7L7B only utilize around 30\% of LUTs and less than 50\% of BRAMs.
This indicates that resource is no longer the bottleneck in ReGraph, benefiting from heterogeneous pipelines customization. 
In addition, with more Little pipelines (hence fewer Big pipelines), LUT and register consumption decrease but BRAM consumption increases.
This is because Little pipelines cost more BRAMs with the Ping-Pong Buffer module, whereas Big pipelines cost more LUTs and registers in the Vertex Loader and Data Router modules. 
Lastly, the frequency is always above 210MHz, benefiting from our crossing SLR optimizations and the efficient resource utilization. 


\subsection{Scalability Exploration}\label{sec:scalability}
Figure~\ref{fig:scalibility} shows the performance of PR with varying the total number of pipelines. 
As one HBM channel only provides 256MB capacity, when the number of HBM channels is small, some graphs are out of memory (marked as `OoM').
The trends in Figure~\ref{fig:scalibility} indicate ReGraph scales well on synthetic graphs or real-world graphs with high average degrees. 
However, super irregular and small graphs are unable to gain linear speedup, which is also observed in previous studies~\cite{chen2021thundergp,isca2021}.
This is because the constant overhead from partition switching overwhelms the speedup of multiple pipelines when partitions are super sparse.

\subsection{Preprocessing Cost}\label{sec:vertexreorderingimpact}
\blue{Table~\ref{table:preprocessingoverhead} shows the preprocessing time of PR on the target CPU with one thread.
Overall, the preprocessing overhead is small and comparable to existing works~\cite{isca2021,chen2021thundergp} as they have the same complexity: $O(E)$ for graph partitioning and $O(V)$ for DBG, where $E$ stands for the number of edges of a graph and $V$ indicates the number of vertices.}

\subsection{Comparison with State-of-the-arts}\label{sec:comparison}
We compare ReGraph on U280 and U50 against ThunderGP~\cite{chen2021thundergp}, Asiatici et al.~\cite{isca2021} and GraphLily~\cite{hu2021graphlily}. 

\noindent
\textbf{Performance.} 
Table~\ref{table:comparefpga} shows the performance comparison between ReGraph and three state-of-the-art works. 
For a more compelling comparison, we ported the open-sourced code of ThunderGP~\cite{thunderGPdoi} to U280. 
It is worth noting that the ported ThunderGP (U280) is 1.3$\times$ faster than original design~\cite{chen2021thundergp}. 
For the other two works, we obtain performance numbers from their papers. 
In short, ReGraph delivers significant speedups to all state-of-the-arts.
Specifically, ReGraph outperforms Asiatici et al.~\cite{isca2021} by up to 5.5$\times$--5.9$\times$, GraphLily~\cite{hu2021graphlily} by 2.1$\times$--3.7$\times$ and ThunderGP (U280) by 1.6$\times$--4.4$\times$. 
Even on U50, a budget platform with only three-quarters of the peak memory bandwidth of U280, ReGraph outperforms GraphLily~\cite{hu2021graphlily} by up to 3.3$\times$ and ThunderGP (U280) by up to 3.7$\times$.

\begin{table}[b!]
\centering
\caption{ReGraph on the U280 and U50 compared to state-of-the-art FPGA-based designs.}
\label{table:comparefpga}
\resizebox{\linewidth}{!}{%
\renewcommand{\arraystretch}{0.75}
\begin{tabular}{@{}cccccc@{}}
\toprule

\multirow{2}[0]{*}{Apps} & \multicolumn{1}{c}{SOTA Works} & \multicolumn{1}{c}{Graph} & \multicolumn{1}{c}{Throughput} & \multicolumn{2}{c}{Our Speedup} \\ 
& (Platform) & Datatsets & (MTEPS) & (U50) & (U280) \\ \midrule

    \multirow{12}[0]{*}{PR} & \multirow{2}[0]{*}{\begin{tabular}[c]{@{}c@{}}Asiatici et al.~\cite{isca2021}\\ (UltraScale+)\end{tabular}} & DB    & 920   & 4.2$\times$   & \textbf{5.9$\times$} \\
          &       & R24   & 1,800  & 4.1$\times$   & \textbf{5.5$\times$} \\ \cmidrule{2-6}
          & \multirow{4}[0]{*}{\begin{tabular}[c]{@{}c@{}}GraphLily~\cite{hu2021graphlily}\\ (U280)\end{tabular}} & R21   & 4,653  & 2.8$\times$   & \textbf{3.3$\times$} \\ 
          &       & HW    & 7,471  & 2.0$\times$   & \textbf{2.1$\times$} \\ 
          &       & PK    & 2,933  & 2.3$\times$   & \textbf{2.8$\times$} \\
          &       & OR    & 5,940  & 1.7$\times$   & \textbf{2.1$\times$} \\  \cmidrule{2-6}
          & \multirow{5}[0]{*}{\begin{tabular}[c]{@{}c@{}}ThunderGP~\cite{chen2021thundergp}\\ (U280)\end{tabular}} & R21   &  5,920     &  2.1$\times$     & \textbf{2.6$\times$} \\ 
          &       & HW    &  6,147     & {2.4$\times$}   &  \textbf{2.5$\times$}\\
          &       & PK    &  3,832     &  1.8$\times$     & \textbf{2.1$\times$} \\ 
          &       & OR    &  5,661     &  2.1$\times$     & \textbf{2.2$\times$} \\ 
          &       & HD    &  1,760     &  4.0$\times$     & \textbf{4.4$\times$} \\ 
                    
          \midrule
          
    \multirow{9}[0]{*}{BFS} & \multirow{3}[0]{*}{\begin{tabular}[c]{@{}c@{}}GraphLily~\cite{hu2021graphlily}\\ (U280)\end{tabular}} & PK    & 1,965  & 3.3$\times$   & \textbf{3.7$\times$} \\
          &       & OR    & 4,937  & 2.3$\times$   & \textbf{2.5$\times$} \\
          &       & HW    & 6,863  & 2.1$\times$   & \textbf{2.2$\times$} \\ \cmidrule{2-6}
          & \multirow{5}[0]{*}{\begin{tabular}[c]{@{}c@{}}ThunderGP~\cite{chen2021thundergp}\\ (U280)\end{tabular}} & R21   &  6,978    &   1.9$\times$    & \textbf{2.0$\times$} \\ 
          &       & HW    &  7,743   & {1.9$\times$}     &  \textbf{1.9$\times$} \\
          &       & PK    &  4,105    &   1.6$\times$    & \textbf{1.8$\times$} \\ 
          &       & OR    &  7,629    &   1.5$\times$    & \textbf{1.6$\times$} \\ 
          &       & HD    &  1,868    &   3.3$\times$    & \textbf{3.7$\times$} \\ 
          
          \midrule
          
    \multirow{5}[0]{*}{CC} & \multirow{5}[0]{*}{\begin{tabular}[c]{@{}c@{}}ThunderGP~\cite{chen2021thundergp}\\ (U280)\end{tabular}} & R21   &  6,182    &   2.1$\times$    & \textbf{2.8$\times$} \\
          &       & HW    &  6,076   & {2.5$\times$}     &  \textbf{3.1$\times$} \\
          &       & PK    &  3,790    &   1.7$\times$    & \textbf{2.0$\times$} \\
          &       & OR    &  5,872    &   2.0$\times$    & \textbf{2.5$\times$} \\
          &       & HD    &  1,737    &   3.7$\times$    & \textbf{4.4$\times$} \\
          \bottomrule

\end{tabular}
}
\end{table}

\begin{figure}[t!]
  \centering
  \includegraphics[width=0.86\linewidth]{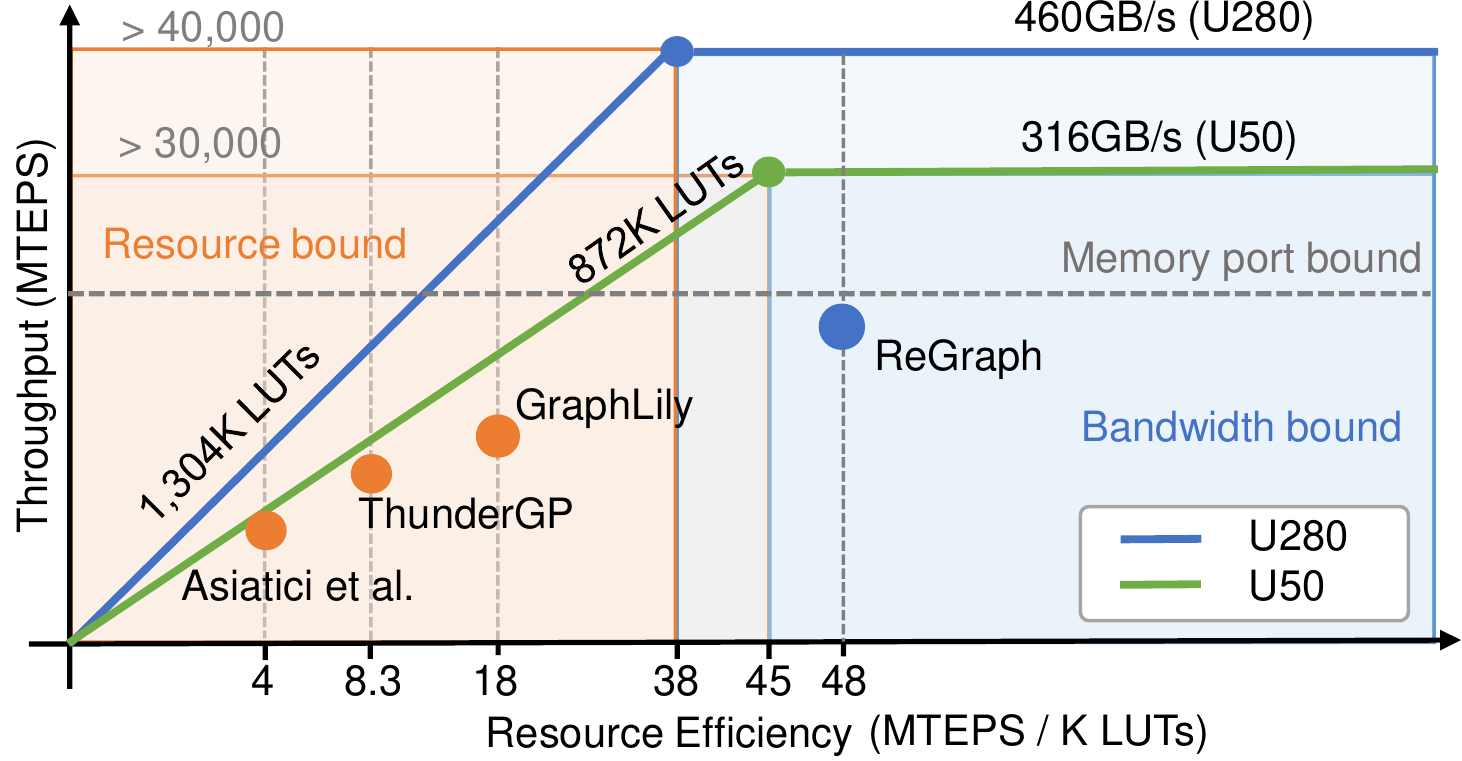}
  \caption{The proposed resource-centric roofline model and resource efficiency comparison with recent designs.}
  \label{fig:roofline}
\end{figure}

\vspace{0.5mm}
\noindent
\textbf{Resource efficiency.}
Figure~\ref{fig:roofline} shows the proposed resource-centric roofline model and resource efficiency comparison. 
While existing {roofline models} primarily focus on operational intensity (flops per memory access)~\cite{siracusa2021comprehensive,diakite2021opencl,williams2009roofline,da2013performance}, ours focuses on resource efficiency (throughput per resource). 
In our roofline model, the x-axis shows the resource efficiency, whereas the y-axis shows absolute performance. 
The horizontal lines represent the bandwidth bounds, while the diagonal lines represent resource bounds. 
We take PR as an example application to calculate the resource efficiency of existing works, by dividing the best reported performance with the reported resource utilization. 
Overall, ReGraph delivers much better resource efficiency than existing works. 
Specifically, ReGraph outperforms Asiatici et al. by 12.3$\times$, ThunderGP by 5.7$\times$ and GraphLily by 2.5$\times$.
While existing works are essentially resource bounded when scaling on U280, ReGraph tackles the resource bottleneck even on U50. 
However, ReGraph is currently bounded by memory ports.

\begin{table}[h!]
\centering
\caption{\blue{{CPU, GPU and FPGA platform specifications. 
Power is measured during execution.}}}
\label{table:cpuconfigurations}
\resizebox{.48\textwidth}{!}{%
\begin{tabular}{@{}lllccc}
\toprule
\multicolumn{1}{l|}{Platform}   & Bandwidth & Power & Process & Release date  \\ \midrule
\multicolumn{1}{l|}{Alveo U280 \textbf{(FPGA)}} & \multicolumn{1}{l}{460 GB/s}  & \multicolumn{1}{c}{35 W} & \multicolumn{1}{c}{16-nm} & \multicolumn{1}{c}{Q4 2018}  \\ \hline
\multicolumn{1}{l|}{Xeon(R) Gold 6248R \textbf{(CPU)}} & \multicolumn{1}{l}{122 GB/s}  & \multicolumn{1}{c}{208 W} & \multicolumn{1}{c}{14-nm} & \multicolumn{1}{c}{
Q1 2020}  \\ \hline
\multicolumn{1}{l|}{Tesla P100 \textbf{(GPU)}}  & \multicolumn{1}{l}{732 GB/s}  & \multicolumn{1}{c}{176 W} & \multicolumn{1}{c}{16-nm} & \multicolumn{1}{c}{
Q2 2016}  \\ \hline
\multicolumn{1}{l|}{Tesla A100 \textbf{(GPU)}}  & \multicolumn{1}{l}{2,039 GB/s}  & \multicolumn{1}{c}{187 W} & \multicolumn{1}{c}{7-nm} & \multicolumn{1}{c}{
Q2 2020}  \\ 
\bottomrule
\end{tabular}%
}
\end{table}

\subsection{Comparison with CPUs and GPUs}\label{sec:gpucomparison}
\blue{
We compare ReGraph on U280 to Ligra~\cite{shun2013ligra} and Gunrock~\cite{wang2016gunrock}, which are the state-of-the-art opensource graph processing frameworks on CPU and GPU, respectively.
Table~\ref{table:cpuconfigurations} shows the configurations of the 48-core CPU platform where we run the latest available Ligra framework~\cite{ligracode} and two different GPU platforms where we run the Gunrock framework~\cite{gunrockcode}.
We measure the CPU power using CPU Energy Meter~\cite{cpumeter}, GPU power using {nvidia-smi} and FPGA power using {xbutil}~\cite{vitis}.
The energy efficiency improvement is calculated as the ratio of ReGraph's GTEPS/Watt to the comparison target's GTEPS/Watt.
}

\begin{figure}[h!]
     \centering
    \begin{minipage}[b]{1\linewidth}
     \includegraphics[width=1\linewidth]{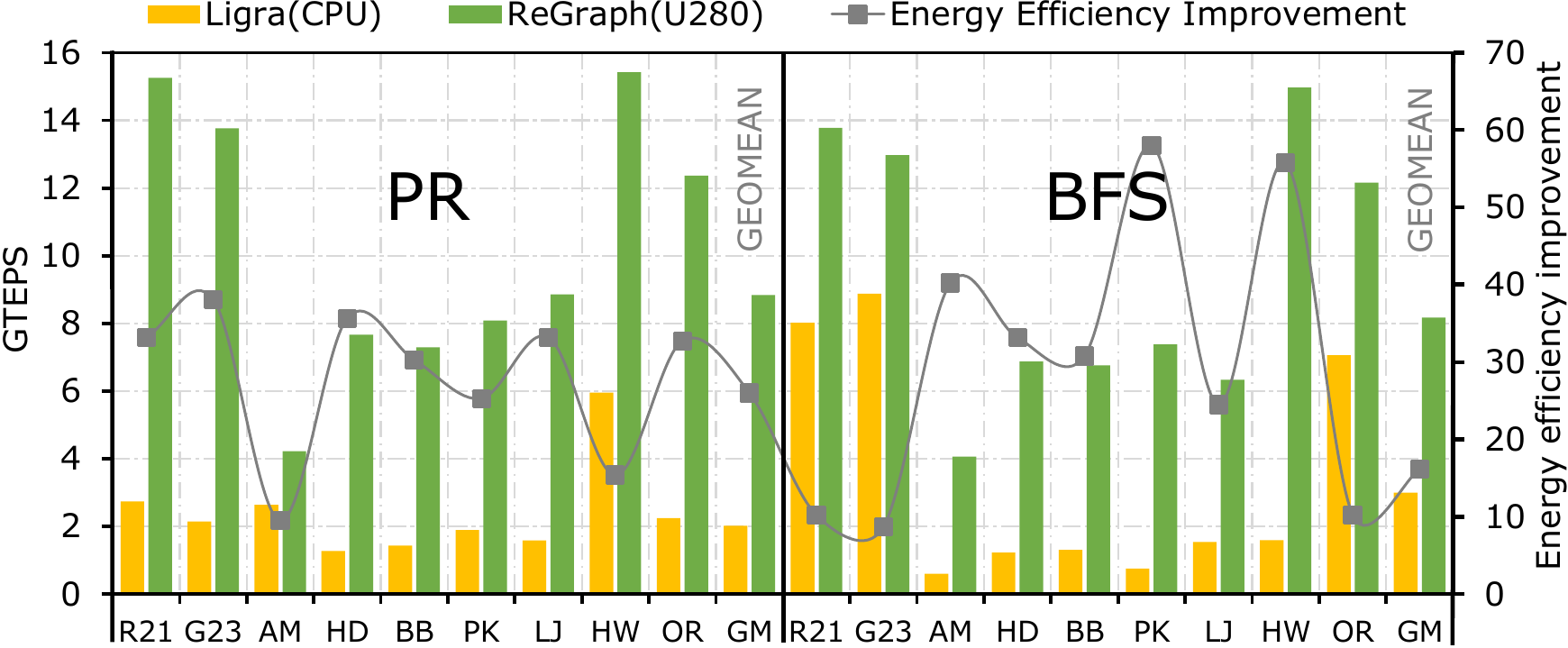}
    \caption{\blue{ReGraph compared to Ligra on CPU.}}
    \label{fig:ligra_pr}
    \end{minipage}
    \par\smallskip 
    \begin{minipage}[b]{1\linewidth}
    \includegraphics[width=1\linewidth]{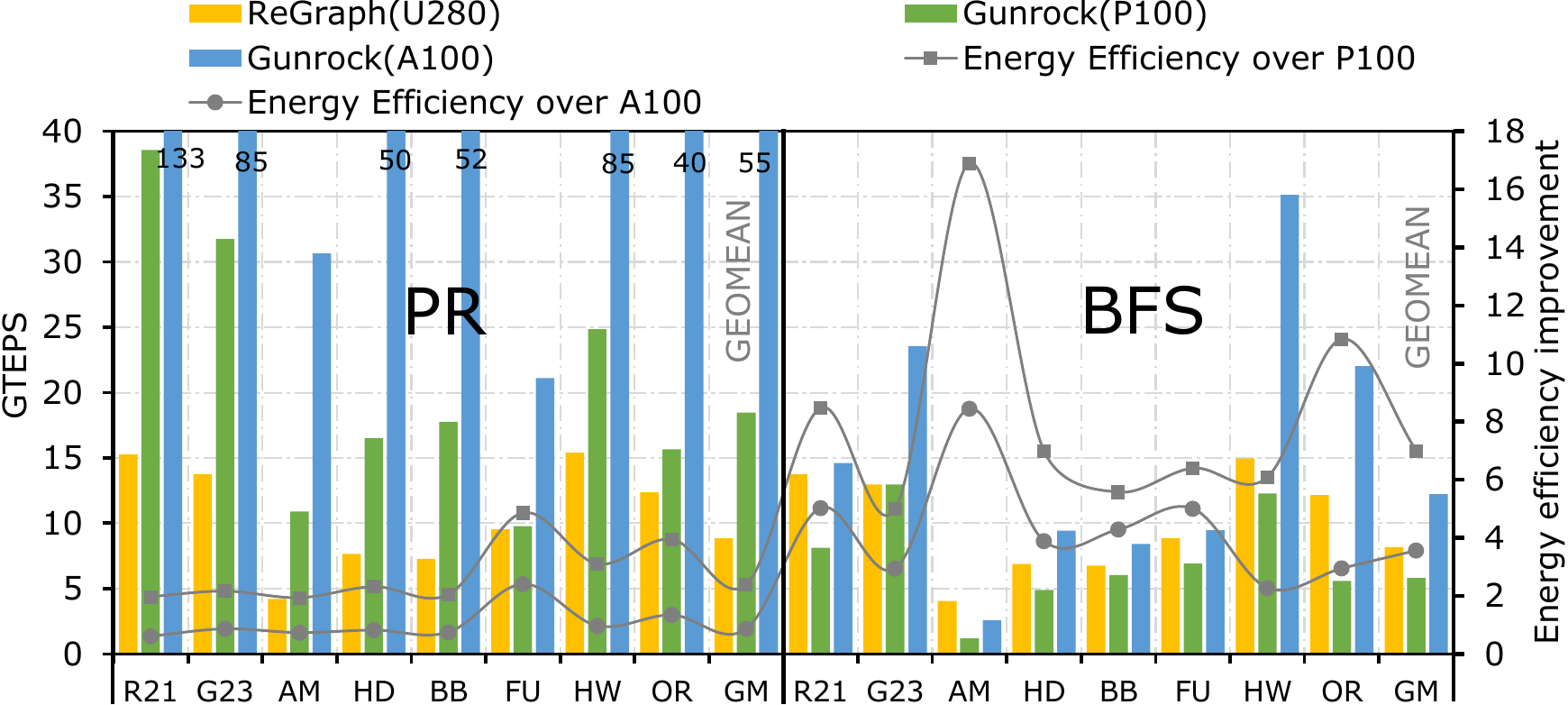}
    \caption{\blue{ReGraph compared to Gunrock on two GPUs.}}
    \label{fig:gunrock}
    \end{minipage}

  \end{figure}

\blue{
Figure~\ref{fig:ligra_pr} shows the comparison between ReGraph and Ligra on a latest server-level CPU.
For PR, ReGraph delivers 1.6$\times$--7.1$\times$ runtime speedup and up to 10$\times$--38$\times$ improvement in energy efficiency. 
For BFS, ReGraph outperforms Ligra by 1.5$\times$--9.7$\times$ in terms of performance and 9.5$\times$--58$\times$ improvement in energy efficiency. 
The significant performance and energy efficiency improvements demonstrate the efficacy of customizing accelerators for graph processing.
}


\blue{
Figure~\ref{fig:gunrock} shows the comparison with Gunrock on Tesla P100 and A100 GPUs. 
For PR, both GPUs perform better than ReGraph in terms of throughput, benefiting from much higher memory bandwidth. 
However, ReGraph delivers 2.4$\times$ (geomean) energy efficiency improvement over P100.
For BFS, ReGraph delivers better performance than P100 and significantly improved energy efficiency: 2.5$\times$--9.2$\times$ improvement (7$\times$ in geomean).
Meanwhile, A100 delivers the best performance with its impressive memory bandwidth and advanced manufacturing process. Still, ReGraph demonstrates an up to 3.5$\times$ (geomean) energy efficiency improvement over A100.
In summary, ReGraph delivers better energy efficiency than GPUs that have the same or even more advanced manufacturing process.
}

\section{Related Work}\label{sec:relatedwork}

In the early stage, ForeGraph~\cite{ForeGraph} explores graph processing with multiple FPGA boards. Later, FabGraph~\cite{ruoshi2019improving} enables two-level vertex buffering technique to ForeGraph and improves performance by 2$\times$. 
Their technique in high-level is overlapping vertex access and edge process, which is similar to our ping-pong buffering design in the Little pipeline. 
However, they only conduct simulation-based experiments, while we implement them efficiently with HLS.
Zhou et al. proposed a series of FPGA-based graph processing works~\cite{Zhou2015pagerank,Zhou2016,Zhou2018CF,zhou2019hitgraph}. 
HitGraph~\cite{zhou2019hitgraph} executes the scatter and the gather stages in a {\em bulk synchronous parallel} (BSP) manner. 
Instead, ReGraph pipelines the Scatter, Gather and Apply stages, hence reducing memory accesses to the global memory. 
Oguntebi et al. presented an open-source modular hardware library, GraphOps~\cite{GraphOps}. 
Chen et al. proposed an OpenCL-based graph processing framework on FPGAs~\cite{chen2019fly}. 
ThunderGP~\cite{chen2021thundergp} fully utilizes the memory bandwidth of the DRAM-FPGA platform. 
Asiatici et al.~\cite{isca2021} proposed to use cache miss optimized memory system for efficient graph processing.
\blue{
However, these solutions suffer from high resource cost, which essentially prevents them from scaling on HBM platforms. 
Although GraphLily~\cite{hu2021graphlily} explored the graph processing on HBM, they failed to customize accelerators as their main technique is to reuse bitstreams of basic modules (e.g., SpMV/SpMSpV).
ReGraph outperforms all the above works significantly in both performance and resource efficiency. 
To the best of our knowledge, we are the first to propose heterogeneous pipeline architectures for graph processing accelerators.}

\blue{
There also exists ASIC-based graph accelerators that demonstrate superior performance under the simulation environment.  
For example, Graphicionado~\cite{ham2016graphicionado} achieves 4.5 GTEPS for PageRank via effective graph partitioning and vertex buffering to an on-chip memory. GraphDynS~\cite{yan2019alleviating} achieves more than 85 GTEPS with HBM (512GB/s) through a hardware/software co-designed approach. 
Ozdal et al. ~\cite{ozdal2016energy} presented an architecture template based on asynchronous execution model to exploit memory-level parallelism, which delivers 3$\times$ speedup over CPU. However, these solutions adopt homogeneous pipeline designs and do not consider resource capacity constraint. 
Rather than having monolithic pipelines, our heterogeneous pipeline designs are lightweight and tailored to diverse workloads of graph processing, delivering significantly improved resource efficiency. 
}


\section{Conclusion and Discussion}\label{sec:conclusion}
HBM-enabled FPGAs have massive memory bandwidth. However, the bottleneck in processing graphs has moved to other resources, making it difficult to fully utilize the bandwidth. In this paper, we propose the use of heterogeneous pipeline architectures to alleviate this issue. 
We first identify two kinds of major workloads within graph processing and showed that the processing of dense vs sparse graph partitions can be optimized in different ways. This gives rise to two customized pipeline types, designed to be
resource-efficient for their specific workloads. 
We also propose an effective task scheduling method that determines pipeline combinations and schedules the graph partitions accordingly. 
\blue{Our framework, ReGraph, further eases the entire development process, delivering up to 5.9$\times$ performance speedup and 12$\times$ resource efficiency improvement compared to the state-of-the-art.}

In our work, we found architectural features that will improve graph processing on future HBM-enabled FPGAs. 
Firstly, logic resources should be increased in general to match the memory level parallelism provided by HBM so that the system is more balanced. 
\blue{Secondly, current HBM restricts graph sizes to smaller than 8~GB. 
As a future work, we plan to introduce SSDs as storage while using HBM as buffers to process billion-scale graphs.}
Thirdly, an increased number of flexible memory ports are needed to improve the utilization of the HBM. ReGraph's performance can be scaled even further once these features are available.





\bibliographystyle{IEEEtranS}
\IEEEtriggeratref{49}
\bibliography{reference}
\end{document}